\documentclass[article,10pt]{elsarticle}
\journal{Mechanical Systems and Signal Processing }
\usepackage[a4paper, total={17cm, 24cm}]{geometry}
\usepackage{hyperref}
\usepackage{lineno,xcolor,graphicx}
\usepackage{caption}
\usepackage[english]{babel}
\usepackage{gensymb}
\usepackage{amssymb,bm}
\usepackage{soul}
\usepackage{mathtools}
\usepackage{amsmath}
\usepackage{expl3}
\expandafter\def\csname ver@l3regex.sty\endcsname{}

\usepackage{ragged2e}

\biboptions{sort&compress}

\begin{document}

\begin{frontmatter}

\title{Analysis of the nonlinear dynamics of a single pendulum driven by a magnetic field using the magnetic charges interaction model and the experimentally fitted interaction model.}
\author[1]{B. Nana\fnref{fn1}}
\ead{na1bo@yahoo.fr}
\author[2]{Krystian Polczy\'{n}ski\fnref{fn2}}
\ead{krystian.polczynski@.p.lodz.pl}
\author[2,3]{P. Woafo\corref{cor1}\fnref{fn3}}
\ead{pwoafo1@yahoo.fr}
\author[2]{Jan Awrejcewicz\fnref{fn4}}
\ead{jan.awrejcewicz@p.lodz.pl}
\author[2]{Grzegorz Wasilewski\fnref{fn5}}
\ead{grzegorz.wasilewski@p.lodz.pl}
\cortext[cor1]{Corresponding author}
\fntext[fn1]{\url{}}
\fntext[fn2]{\url{https://orcid.org/0000-0002-1177-6109}}
\fntext[fn3]{\url{https://orcid.org/0000-0002-7918-4118}}
\fntext[fn4]{\url{https://orcid.org/0000-0003-0387-921X}}
\fntext[fn5]{\url{https://orcid.org/0000-0002-5549-2976}}
\address[1]{Department of Physics, Higher Teacher Training College, University of Bamenda,
	PO Box 39 Bamenda, Cameroon}
\address[2]{Lodz University of Technology, Department of Automation, Biomechanics, and Mechatronics, 1/15
	Stefanowskiego Str., 90-537 Lodz, Poland}
\address[3]{Laboratory of Modelling and Simulation in Engineering, Biomimetics and Prototypes,
Faculty of Science, University of Yaounde I, PO Box 812 Yaounde, Cameroon}

\begin{abstract}
In this work, we analyzed theoretically and experimentally the nonlinear dynamics of a magnetic pendulum driven by a coil-magnet interaction.  The force between the magnetic elements and the resulting torque on the pendulum are derived using both the magnetic charges interaction model and the experimentally fitted interaction model. This enables the comparison between the two models. The current in the coil is taken first as a sinusoidal current and then as a square current. The comparison of the structure of each interaction model is conducted and it appears that they give qualitatively similar characteristics. The harmonic balance method is used to approximate the frequency responses of the pendulum leading to both symmetric and asymmetric or one-side (intrawell) oscillations. The two-parameters bifurcation diagrams are plotted showing the different dynamical behaviors considering the current amplitude and frequency as the control parameters. Good agreements are found between our theoretical results and experimental ones.
\end{abstract}


\end{frontmatter}
\section{Introduction}

$\quad$ In recent years pendulum systems are extensively analyzed due to a variety of reasons. In fact, pendulum systems are the source of motivation for many discoveries in nonlinear dynamics and other engineering domains \cite{aa0,aa1,aa2,aa3,aa4,bb0,bb1,bb2,bb3,bb4,bb5}. Pendulum-like systems have been also used to model biological systems and electrical motors \cite{bb6,bb7}. Results show several complex responses that include bifurcations, chaos, multistability, transient chaos, and hysteresis. Contrary to electronic and electrical circuits, where one has to monitor chaotic behavior on the oscilloscope, pendulum systems allow direct observation of the chaotic motion of the bobs or masses of the system.

Being involved in the development of a wave energy converter, the parametrically excited pendulums have significant interest. Parametric excitation is a phenomenon widely encountered in electrical and mechanical dynamical systems. The fundamental characteristics of such systems are that the stiffness coefficients and in some cases the damping parameters are functions of the external excitation. The most common example is a simple pendulum whose pivot executes vertical periodic oscillations \cite{bb9}. The linear differential equation known as the Mathieu equation that describes the dynamics of such systems has been shown to give rise to unstable response when resonating \cite{bb10}.

Among the parametric systems analyzed in the literature, magnetic pendulums are of higher interest given their numerous applications in engineering. \textcolor{black}{For example, in Ref. \cite{bKol200} was investigated a novel autoparametric pendulum absorber utilizing magnetic forces for enhanced vibration damping. Analytical analyses of axial and lateral magnetic forces, validated numerically, preceded an assessment of the primary system's response with the new absorber. Results indicated superior performance compared to conventional absorbers, with reduced vibration amplitudes and an expanded absorption bandwidth. Key parameters for optimal absorber design were highlighted. Subsequent research involved the experimental validation of this innovative absorber design.} Recent works have also considered pendulum setups in external magnetic fields where ferrofluids are used either as surrounding medium or as magnetic material of the bob itself \cite{bb11,bb12}. The main nonlinearity in such systems is due to the interaction between permanent magnets \cite{bb13,bb14} or the interaction between permanent magnets and electric coils \cite{bb15,b1,bKol15}.
Witkowski et al. \cite{bKol61} conducted an investigation into a two-degree-of-freedom system characterized by magnetic interaction. The system comprises two mechanically coupled carts, whose movement is constrained by nonlinear magnetic springs. The researchers performed a comprehensive comparison of various mathematical models of the system and their parameters through both numerical simulations and experimental studies. The dynamic behavior of the system was revealed to encompass a chaotic attractor crisis, quasi-periodic solutions, and finite periodic doubling cascades.
In Ref.\cite{b1}, the authors approximated the mathematical expression of interactions between a permanent magnet and an electric coil in a pendulum system using data obtained experimentally. They found an excellent agreement between their theoretical and experimental results. Parallel to this experimentally fitted coil-magnet interaction, there is another model which considers magnet and coil as charges and uses an equivalent of Coulomb law to derive the expressions of the force between them \cite{bb13}. This model has also lead to interesting dynamics for magnetic pendulum (theoretically and experimentally) \cite{bb14}.

\textcolor{black}{
The nonlinear dynamics of a pair of coupled pendulums was explored in \cite{bKol52,bKol53,bKol54}. In this study, one pendulum was subjected to a magnetic force, while the second one was set in motion through torsion coupling facilitated by a flexible element. The researchers developed a comprehensive mathematical model for the system, and the formula for the magnetic torque was derived based on experimental data. Through extensive bifurcation analysis and Poincaré sections, a diverse range of behaviors emerged, including chaotic, multi-periodic, and quasi-periodic solutions.
The investigation revealed symmetrical structures in the obtained basins of attraction for various periodic and quasi-periodic solutions. The presence of quasi-periodicity was further elucidated through the calculation of Lyapunov exponents and Fourier spectrums. Importantly, both numerical simulations and experimental studies demonstrated a high level of agreement, underscoring the robustness and reliability of the findings.}

\textcolor{black}{
The chaotic and regular behavior of a system comprising a double physical pendulum with two repulsive permanent magnets was explored in \cite{bKol58}. The focus was on mathematical modeling, numerical simulations, experimental measurements, with particular emphasis on a novel magnetic interaction model. System parameters were identified by aligning output signals from experiments and numerical solutions with the developed mathematical model governed by a strongly nonlinear set of two second-order ordinary differential equations, encompassing friction and magnetic interaction torques.
The considered system exhibited both chaotic and periodic dynamics. Several numerically identified chaotic zones were confirmed experimentally. Scenarios depicting the transition from regular to chaotic motion and vice versa, along with bifurcation diagrams, were illustrated and discussed. The study achieved good agreement between numerical simulations and experimental measurements.
The intricate dynamics exhibited by a system of double pendulums with magnets sparked inspiration for Kumar et al. \cite{bKol60} to explore its potential as an energy harvester. Through both numerical simulations and experimental analyses, the researchers found that the harvested power experienced a substantial increase when the system's dynamics entered a chaotic motion phase.
Malaji et al. \cite{bKol90} also experimentally and numerically investigated the possibilities of energy harvesting in a system of two magnetic pendulums but coupled in parallel.
Sosna et al. \cite{bKol100} studied a vibrating beam (which plays the role of a pendulum)  with magnet at the tip to explore advanced methods for converting vibrational kinetic energy into usable electricity, with a focus on piezoelectricity. Results indicated that, for a vibration amplitude of 0.5 g, the monostable regime was suitable for frequencies lower than the linear resonator's natural frequency. In the bistable regime, the most effective oscillation type was in-well single-periodic behavior, suitable for tuning across the entire frequency range from 20 to 65 Hz. 
More studies on multistable systems using magnetic interactions in relation to energy harvesting can be found in the review research preformed by Fang et al. \cite{bKol80}.}

\textcolor{black}{
Pilipchuk et al. \cite{bKol55} introduced a methodology aimed at controlling resonance energy exchange within a mechanical system comprising two weakly coupled magnetic pendulums interacting with a magnetic field generated by coils positioned underneath. The paper illustrates that appropriately guided magnetic fields can effectively alter mechanical potentials, thereby directing energy flow between the pendulums in a desired manner. In antiphase oscillations, energy transfers from the pendulum exposed to the repelling magnetic field to the oscillator influenced by the attracting magnetic field. Conversely, during inphase oscillations, the energy flow is reversed.
The closed-loop controller, relying solely on information about the phase shift, can be estimated from dynamic state signals through the coherency index. The authors emphasize the advantageous nature of the proposed control strategy, operating at a relatively slow temporal rate compared to the inherent oscillations of the system.}

\textcolor{black}{
Magnetic interactions between magnets and electric conductors also have good damping properties, which are presented in paper \cite{bKol15pop}. This study investigated the design and dynamics of a rotary eddy current damper (ECD). Utilizing analytical methods and finite element simulations, the eddy current torque was evaluated across varying conductor speeds. Experimental validation was conducted for both a simple torque generator and a full-scale rotary ECD. Results indicated consistency between analytical and simulated torques, though they were higher than experimental values. Importantly, a temperature rise in the large-scale ECD impacted damping performance. Accounting for temperature effects, numerically predicted damping aligned well with experiments, demonstrating performance comparable to fluid viscous dampers. The rotary ECD exhibited highly nonlinear damping forces, influencing its effectiveness at different speeds. In seismic control applications, both fluid viscous dampers and rotary ECDs effectively reduced peak and RMS displacement responses in different structural systems.
Similar behaviors were discovered in the magnetic pendulum system in paper \cite{bKol49}.
The problem was transformed into the Mathieu equation, and the harmonic balance method was employed to investigate the conditions for instability of its equilibrium position with electromagnetic interaction. The paper additionally presented numerical analysis results, which encompassed the emergence of doubly connected areas of harmonic instability, the coexistence of stable periodic orbits, and chaotic motions during the subharmonic instability under moderately strong driving.}

\textcolor{black}{
In this work, we considered two models of the magnetic pendulum system. The first model used a theoretically derived mathematical model of a magnetic interaction, while the second model used a magnetic interaction model based on empirical studies. 
The analysis was aimed to check the correctness of both models based on the conducted studies of  the system dynamics and to compare the obtained results.
The system consists of a magnetic pendulum coupled to a current-carrying coil. The permanent magnet is positioned at the lower end of the pendulum and the coil is directly below the pendulum vertical position. 
This system possesses two control parameters: the frequency and the amplitude of the current source supplying the coil. We analyzed changes in the system potential depending on the current value, showing that for a positive current, the system has a double-well potential, while for a negative current, the system has a single-well potential with a significantly deepened well. Moreover, we analyzed analytically and numerically, the influence of the coil current value on changes in the pendulum's lower equilibrium position and its impact on the stability of these equilibriums. Two-dimensional bifurcation diagrams were calculated, demonstrating the periodic and chaotic nature of the magnetic pendulum and the impact of changing these parameters on the dynamics of the system.} 

The paper is organized as follows. The physical description and the mathematical models of the system are presented in Section 2. The mathematical development and the numerical simulations are carried out in Section 3 and 4, respectively. The comparison between our theoretical and experimental results is presented in Section 5. Finally, the conclusion and perspectives are given in the last section.

\section{Physical description and mathematical model}

\subsection{Physical description}
\begin{figure}[h!]
\begin{center}
\includegraphics[width=7.5cm,height=6.0cm]{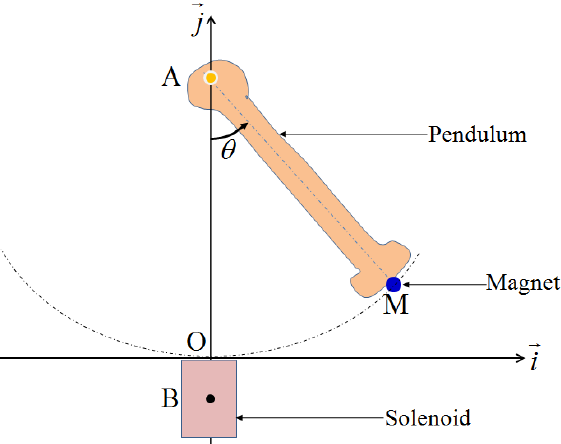}\hspace*{0cm}
\caption{\label{FIGURE1}\footnotesize {Magnetic pendulum in a variable magnetic field created by a current-carrying coil.}}
\end{center}
\end{figure}

$\quad$ A schematic representation of the studied system driven by a magnetic force is shown in Figure \ref{FIGURE1}, where $\theta$ is the angular deviation.

As indicated in the figure, a strong neodymium magnet is joined to the free end of the pendulum, while the upper end of the pendulum is fixed to a horizontal rotational axis. A current-carrying coil is placed under the pendulum arm, and the axis of the coil coincides with the vertical axis, namely the $y-$axis. Point B is the center of the coil such that $OB=r$ and $\ell$ represents the length of the pendulum.

\subsection{Mathematical models}

$\quad$ For simplicity of our theoretical model, we assume that the magnetic force between the coil and the magnet is an inverse squared dependence on distance, and it is thus a function of the angular displacement of the pendulum $\theta$. Considering magnetic charges $Q_1$ and $Q$ for the magnet and the coil, respectively. The magnetic force $\overrightarrow{F}$ between the coil and the magnet can be written as
\begin{equation}\label{EQUAT1}
	\overrightarrow F  = \frac{{\mu _0 Q_1 Q}}{{4\pi }}\frac{{\ell \sin \theta \overrightarrow i  + \left( {\ell  + r - \ell \cos \theta } \right)\overrightarrow j }}{{\left( {2\ell ^2  + 2\ell r + r^2  - 2\ell \left( {\ell  + r} \right)\cos \theta } \right)^{\frac{3}{2}} }}.
\end{equation}

Here, $\mu_0$ is the permeability of the vacuum. Let us mention that this formulation assumes that the magnetic field is the same in all directions (which can be a rough assumption for some magnet shapes). Since the magnetic field provided by the coil is due to the current $i(t)$ flowing through the windings, the magnetic charge $Q_1$ is proportional to the current and can be written as $Q_1=ki(t)$. The magnetic torque produced by the magnetic force $\overrightarrow{F}$ has the following expression:
\begin{equation}\label{EQUAT2}
	\Gamma_1  =\frac{{kQ\mu _{\text{0}} }}{{4\pi }} \frac{{i\left( t \right)\ell \left( {\ell  + r} \right)\sin \theta }}{{\left( {2\ell ^2  + 2\ell r + r^2  - 2\ell \left( {\ell  + r} \right)\cos \theta } \right)^{\frac{3}{2}} }}.
\end{equation}
After some mathematical transformations, equation (\ref{EQUAT2}) can be simplified as
\begin{equation}\label{EQUAT3}
	\Gamma _1  = \frac{{\sigma i\left( t \right)\sin \theta }}{{\left( {1+\varepsilon - \cos \theta } \right)^{\frac{3}
				{2}} }},\;\;{\text{where}}\;\sigma  = \frac{{k\mu _0 Q}}{{8\pi \sqrt {2\ell \left( {\ell  + r} \right)} }}\;{\text{and}}\;\varepsilon  =\frac{{r^2 }}{{2\ell \left( {\ell  + r} \right)}}.
\end{equation}
In this work, the current $i(t)$ is provided by a current source. Hence, the effect of the induced electromotive force due to the movement of the neodymium magnet in the coil is neglected. The application of Newton's second law to the pendulum leads to the equation (\ref{EQUAT4})
\begin{equation}\label{EQUAT4}
	J\ddot \theta  + \beta \dot \theta + C\theta  + mg\ell_1 \sin \theta + \Gamma _s \left( {\dot \theta } \right) = \Gamma_1.
\end{equation}
Terms $J$ and $m$ are a moment of inertia and a mass of the pendulum with the fixed magnet, respectively.
 Parameter $\beta$ is a complete viscous damping coefficient and parameter $C$  is the stiffness of the joint. Parameter $\ell_1$ represents the distance from the pivot to the center of mass of the magnetic pendulum, while the parameter $g$ is the gravitational constant. Finally, term $\Gamma _s \left( {\dot \theta } \right)$ represents the nonlinear part of the Stribeck friction torque and is expressed as
\begin{equation}\label{EQUAT7}
	\Gamma _s \left( {\dot \theta } \right) = \left[ {M_c  + \left( {M_s  - M_c } \right)\exp \left( { - \frac{{\dot \theta ^2 }}{{v_s }}} \right)} \right]\tanh \left( {\chi \dot \theta } \right).
\end{equation}
In the above equations, $M_c$ represents the magnitude of Coulomb friction, $M_s$ is the static friction value and the parameter $v_s$ stands for Stribeck velocity. Based on the experimental measurements, the mathematical model of this system has been proposed by the authors of reference \cite{b1} and can be summarized as follows
\begin{equation}\label{EQUAT5}
	J\ddot \theta  + \beta \dot \theta  + C\theta  + mg\ell_1 \sin \theta  + \Gamma _s \left( {\dot \theta } \right) = \Gamma _2,
\end{equation}
where $\Gamma_2$ is the torque exerted by the coil on the neodymium magnet. They found during their experimental investigations that $\Gamma_2$ is proportional to the current through the coil and is related to the mechanical displacement $\theta$ as shown in equation (\ref{EQUAT6})
\begin{equation}\label{EQUAT6}
	\Gamma _2  = \frac{{2i\left( t \right)a\theta }} {b}\exp \left( { - \frac{{\theta ^2 }}{b}} \right).
\end{equation}
The coefficients $a$ and $b$ are fitting parameters. These fitting parameters are also functions of the amplitude of the current through the coil.

\subsection{Comparison between the models}

$\quad$ The only difference between the two models described by equations (\ref{EQUAT4}) and (\ref{EQUAT5}) lies in the expression of the magnetic torques $\Gamma_1$ and $\Gamma_2$. Let us remind that the magnetic charges can be measured using a dynamometer given the forces between magnetic elements versus the distance separating the elements or using a pendulum experiment in which the tangent of the angle made by the pendulum is equal to the ratio between the magnetic force between the magnets and the weight of the magnet fixed at the end of the pendulum.

\begin{figure}[h!]
	\begin{center}
		\includegraphics[width=6.3cm,height=4.7cm]{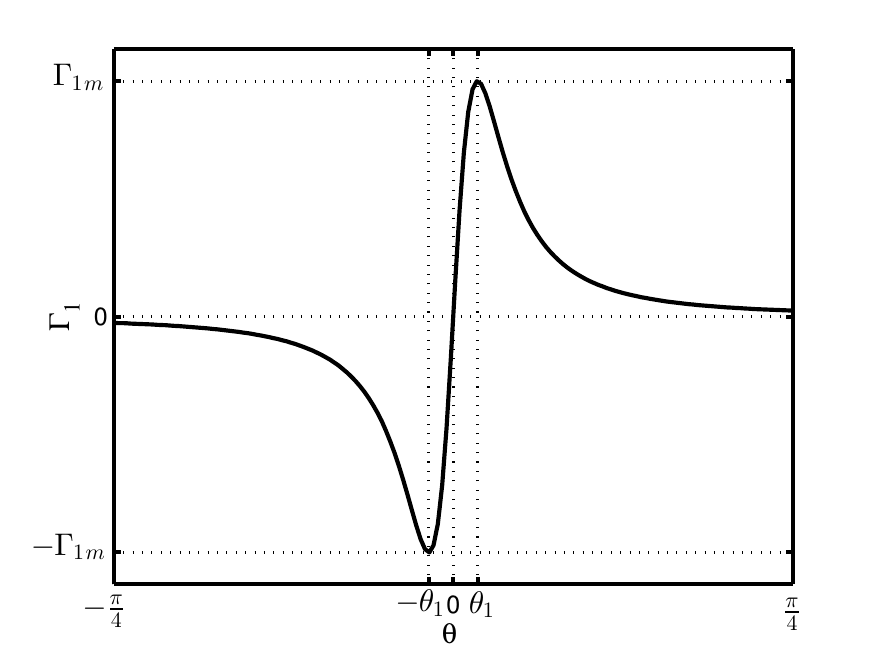}\hspace*{0.0cm}
		\includegraphics[width=6.3cm,height=4.7cm]{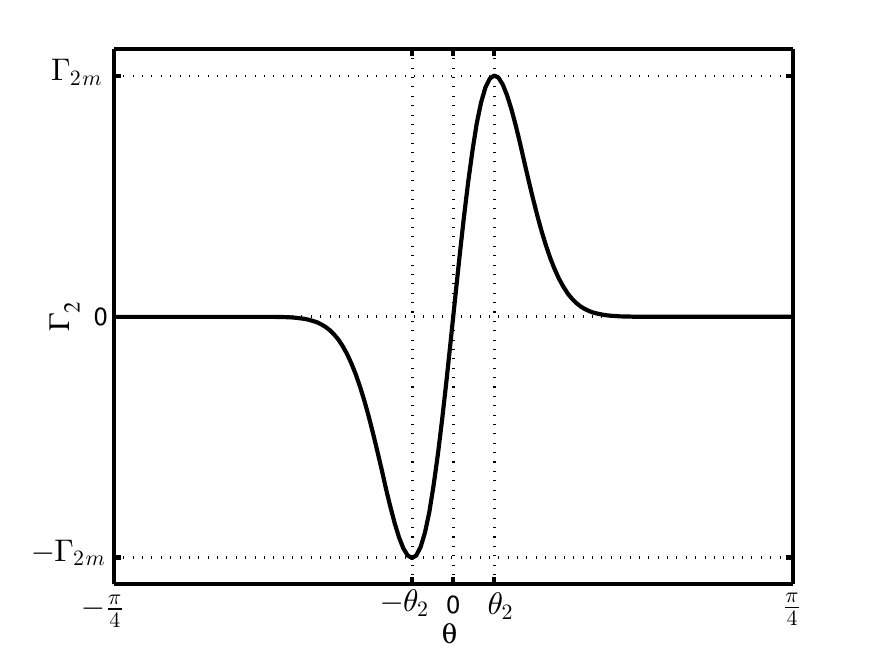}\\
		\hspace*{1.2cm} a) \hspace*{6.0cm} b) \hfill\\
		\caption{\label{FIGURE2}\footnotesize {(a) Shape of the torque $\Gamma_1$ and (b) shape of the torque $\Gamma_2$, both obtained for $1$ A of the coil current.}}
	\end{center}
\end{figure}

Both models show that the torque exerted on the pendulum by the coil is significant for a certain range of the state variable $\theta$ called the active zone. For illustration, Figure \ref{FIGURE2} shows the shapes of the torques $\Gamma_1$ and $\Gamma_2$ for a constant coil current.
Even if the limit of the active zone is smooth in Figure \ref{FIGURE2}a, the two curves have approximatively the same shape. The parameters labeled on the graphs are defined as
\begin{equation}\label{EQUAT8a}
	\theta _1  = \arccos \left( {\sqrt {\varepsilon ^2  + 2\varepsilon  + 4} - \varepsilon  - 1} \right),\;\;\theta _2  = \sqrt {\frac{b}{2}},
\end{equation}
\begin{equation}\label{EQUAT8b}
	\Gamma _{1m}  = \sigma i\left( t \right)\sqrt {\frac{{1 - \left( {\sqrt {\varepsilon ^2  + 2\varepsilon  + 4}  - \varepsilon  - 1} \right)^2 }}{{\left( {2\varepsilon  + 2 - \sqrt {\varepsilon ^2  + 2\varepsilon  + 4} } \right)^3 }}}, \;\;\Gamma _{2m}  = ai\left( t \right)\sqrt {\frac{2}{{be}}}.
\end{equation}
One can notice that the first torque $\Gamma_1$ is described by two parameters ($\varepsilon$, $\sigma$) and the second torque $\Gamma_2$ is also described by two parameters ($a$, $b$). For the two torques to be quantitatively very closed, the mathematical relations have to be defined between these four parameters. The comparison of $\theta_1$ and $\theta_2$ in equation (\ref{EQUAT8a}) gives $\varepsilon$ as a function of $b$. On the other hand, the comparison of $\Gamma _{1m}$ and $\Gamma _{2m}$ in equation (\ref{EQUAT8b}) gives $\sigma$ as a function of the parameters $a$ and $b$.

From the experimental values obtained by the authors of reference \cite{b1}, the values of $\varepsilon$ and $\sigma$ have been computed and the values of the parameters $r$ and $Q$ have been deduced. The values of the parameters used in this work are given in Table \ref{CH1T1}.\\
\begin{table}[h]
	\begin{center}
		\begin{tabular}{|c|l||c|l|}\hline
			Parameters	&  Values &	Parameters	&  Values\\\hline
			$a$ & $4.253 \cdot 10^{ - 2} \;{\text{Nm}}$ & $b$ & $1.818 \cdot 10^{ - 2} \;{\text{rad}}$\\
            $Q$ & $40.0  \;{\text{Am}}$ & $k$ & $1425 \;{\text{m}}$\\
			$r$ & $0.100$ m & $\ell$ & $1.200$ m\\
			$J$ & $6.787 \cdot 10^{ - 4} \;{\text{kgm}}^{\text{2}}$ & $\gamma  = mg\ell _1$ & $5.800 \cdot 10^{ - 2} \;{\text{Nm}}$\\
			$\beta$ & $2.019 \cdot 10^{ - 4} \;{\text{Nm/rad}}$ & $C$ & $1.742 \cdot 10^{ - 2} \;{\text{Nm/rad}}$\\
			$M_c$ & $2.223 \cdot 10^{ - 4} \;{\text{Nm}}$ & $M_s$ & $4.436 \cdot 10^{ - 4} \;{\text{Nm}}$\\
			$v_s$ & $5.374 \cdot 10^{ - 1} \;{\text{rad/s}}$ & $\chi$ & $5.759 $\\
			$f$ & $0.2{\text{ Hz}} \leqslant f \leqslant 10.0{\text{ Hz}}$ & $I_m$ & $0.0{\text{ A}} \leqslant I_m  \leqslant 5{\text{ A}}$ \\\hline
		\end{tabular}
		\caption{Values of the parameters used in this work \cite{b1}.}\label{CH1T1}
	\end{center}
\end{table}

\subsection{Equilibrium points and stability analysis}

\begin{figure}[h!]
	\begin{center}
		\includegraphics[width=6.3cm,height=4.5cm]{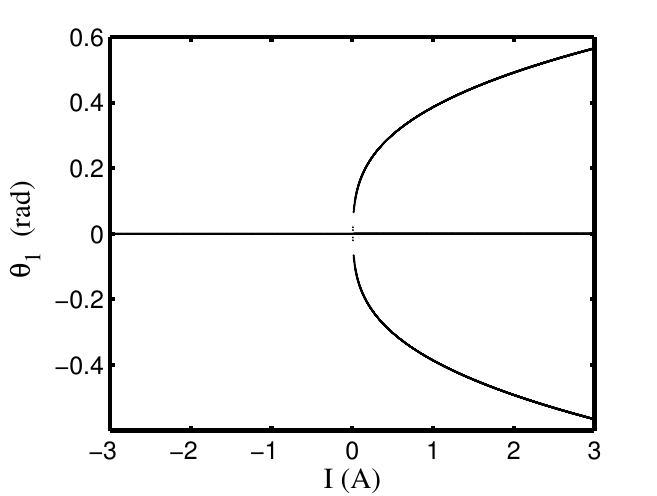}\hspace*{0.0cm}
		\includegraphics[width=6.3cm,height=4.5cm]{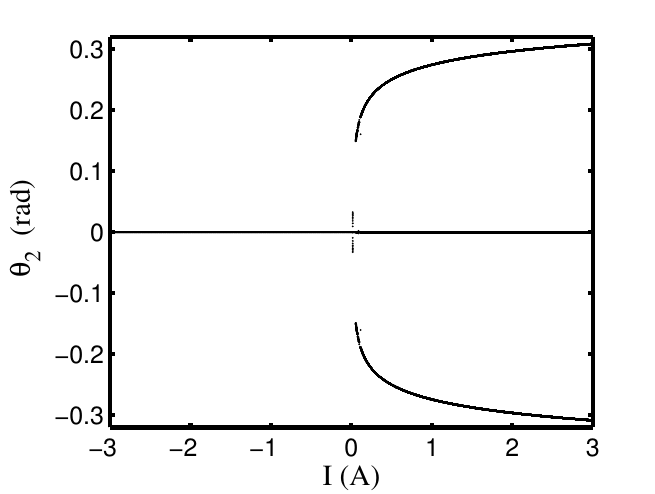}\\
		\hspace*{0.3cm} a) \hspace*{5.5cm} b) \hfill\\
		\caption{\label{FIGURE3}\footnotesize {
				Equilibrium points of the system obtained using the ﬁrst model (a) and the second model (b).}}
	\end{center}
\end{figure}

$\quad$ In this analysis, the time-dependent current source $i(t)$ is replaced by a constant current source which provides a current $I$. For the first and second models described above, the equilibrium points are, respectively, the roots of equations (\ref{EQUAT9}) and (\ref{EQUAT10}):
\begin{equation}\label{EQUAT9}
	\left( {C\theta  + \gamma \sin \theta } \right)\left( {1 + \varepsilon  - \cos \theta } \right)^{\frac{3}
		{2}}  - \sigma I\sin \theta  = 0,
\end{equation}
\begin{equation}\label{EQUAT10}
	b\left( {C\theta  + \gamma \sin \theta } \right) = 2aI\theta \exp \left( { - \frac{{\theta ^2 }}
		{b}} \right).
\end{equation}
\textcolor{black}{The roots of the above equations are not directly accessible. To obtain their precise roots, we will find the series expansion of the equations. The greater the degree of the polynomial, the more precise the obtained roots will be. We will therefore limit ourselves to a degree for which the analytical analysis is possible.} Taking the $6^{{\text{th}}}$ order series expansion of the equation (\ref{EQUAT9}) leads to the following polynomial equation
\begin{equation}\label{EQUAT11}
	\begin{gathered}
		A_0 \theta  + 2A_1 \theta ^3  + A_2 \theta ^5  = 0,\;{\text{where}} \hfill \\  A_0  = 480\left( {C\varepsilon ^2  + \gamma \varepsilon ^2  - \sigma I\sqrt \varepsilon  } \right),\hfill \\A_1  = 20\left( {9C\varepsilon  + 9\gamma \varepsilon  - 2\gamma \varepsilon ^2  + 2\sigma I\sqrt \varepsilon  } \right), \hfill \\
		A_2  = 45C - 30C\varepsilon  + 45\gamma  - 90\gamma \varepsilon  + 4\gamma \varepsilon ^2  - 4\sigma I\sqrt \varepsilon . \hfill \\\end{gathered}
\end{equation}
On the other hand, the logarithm function is first applied to both sides of the equation (\ref{EQUAT10}) before making the $6^{{\text{th}}}$ order series expansion, it comes that
\begin{equation}\label{EQUAT12}
	\begin{gathered}
		B_0 \theta  + 2B_1 \theta ^3  + B_2 \theta ^5  = 0,\;{\text{where}} \hfill \\  B_0  = -360\ln \left( {\frac{{2aI}}{{b\left( {C + \gamma } \right)}}} \right),\;B_1  = 30\frac{{6C + 6\gamma  - b\gamma }}
		{{b\left( {C + \gamma } \right)}}\;{\text{and}}\;B_2  = \frac{{\gamma \left( {3C - 2\gamma } \right)}}
		{{\left( {C + \gamma } \right)^2 }} \hfill \\\end{gathered}
\end{equation}
The number of equilibrium points of the system depends on the sign of the product $IQ$. Assuming a positive value of $Q$, if $I\leq 0$, $\theta=0$ is the unique equilibrium point as the only real root of equations (\ref{EQUAT11}) and (\ref{EQUAT12}). If $I>0$, the above polynomial equations have three real roots given respectively as
\begin{equation}\label{EQUAT13}
	\theta _{11}  =  - \sqrt {\frac{{\sqrt {A_1^2  - A_0 A_2 }  - A_1 }}{{A_2 }}} ,\;\theta _{12}  = 0\;{\text{and}}\;\theta _{13}  = \sqrt {\frac{{\sqrt {A_1^2  - A_0 A_2 }  - A_1 }}{{A_2 }}} .
\end{equation}
\begin{equation}\label{EQUAT14}
	\theta _{21}  =  - \sqrt {\frac{{\sqrt {B_1^2  - B_0 B_2 }  - B_1 }}{{B_2 }}} ,\;\theta _{22}  = 0\;{\text{and}}\;\theta _{23}  = \sqrt {\frac{{\sqrt {B_1^2  - B_0 B_2 }  - B_1 }}{{B_2 }}} .
\end{equation}
To verify these results, the roots of equations (\ref{EQUAT9}) and (\ref{EQUAT10}) are plotted respectively in Figures \ref{FIGURE3}a,b using the Newton-Raphson algorithm.

\begin{figure}[h!]
	\begin{center}
		\includegraphics[width=6.3cm,height=4.5cm]{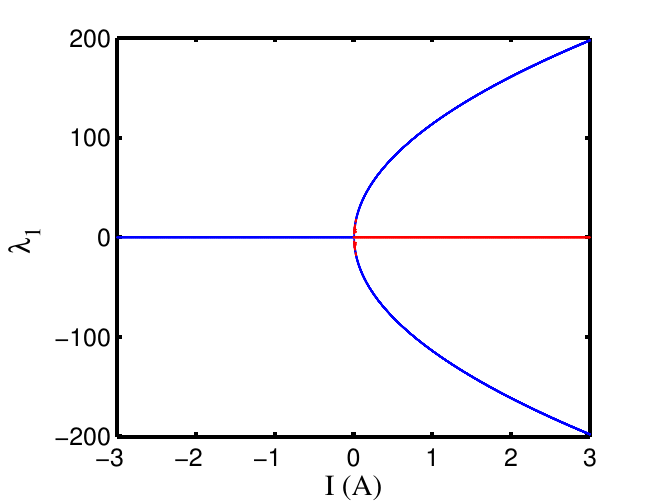}\hspace*{0.0cm}
		\includegraphics[width=6.3cm,height=4.5cm]{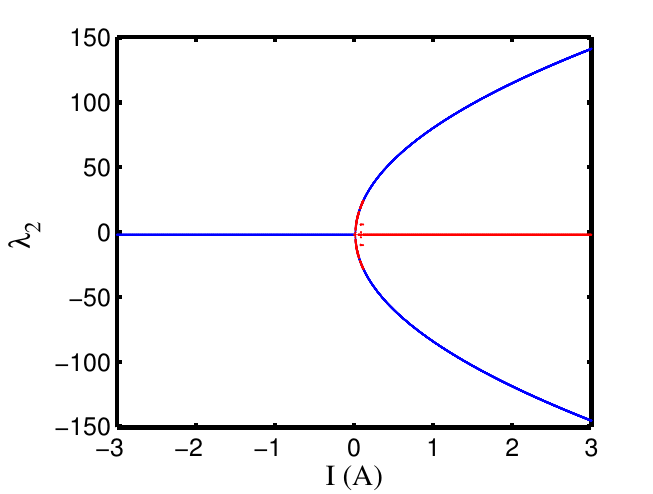}\\
		\hspace*{1.2cm} a) \hspace*{6.0cm} b) \hfill\\
		\caption{\label{FIGURE4}\footnotesize {
				Nature of the equilibrium points calculated using the first model (a) and the second model (b). Blue color represents a trivial root $\theta=0$, while red color represents the real part of the opposite roots.}}
	\end{center}
\end{figure}

As mentioned previously and verified by the curves, the system has one equilibrium point for negative current and three equilibrium points for positive current. One can notice the qualitative similarity between the results obtained from the two models. To determine the nature of the equilibrium points, the real parts of the corresponding eigenvalues are plotted in Figure \ref{FIGURE4}a for the first model and in Figure \ref{FIGURE4}b for the second model.
In both figures, the curves in blue represent the real part of the trivial root $\theta=0$, and the curves in red indicate the real part of the two opposite roots. As the graphs reveal, the equilibrium point $\theta=0$ is stable for $I<0$ and becomes unstable for $I>0$ while the other equilibrium points are stable.

\begin{figure}[h!]
	\begin{center}
		\includegraphics[width=6.3cm,height=5.0cm]{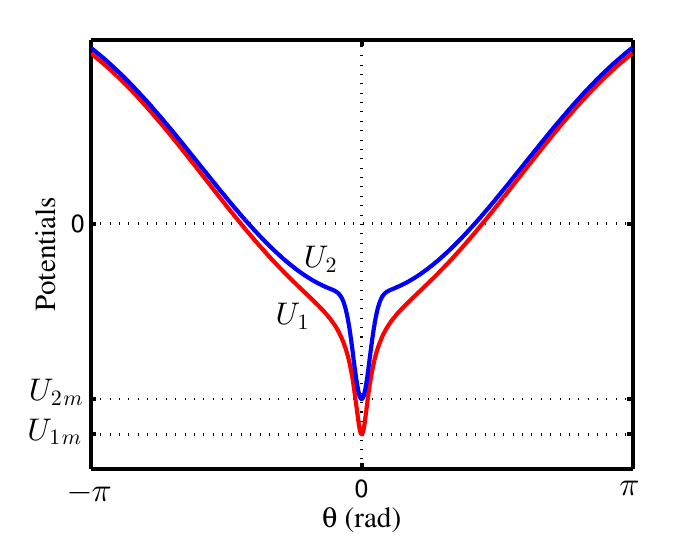}\hspace*{0.0cm}
		\includegraphics[width=6.3cm,height=5.0cm]{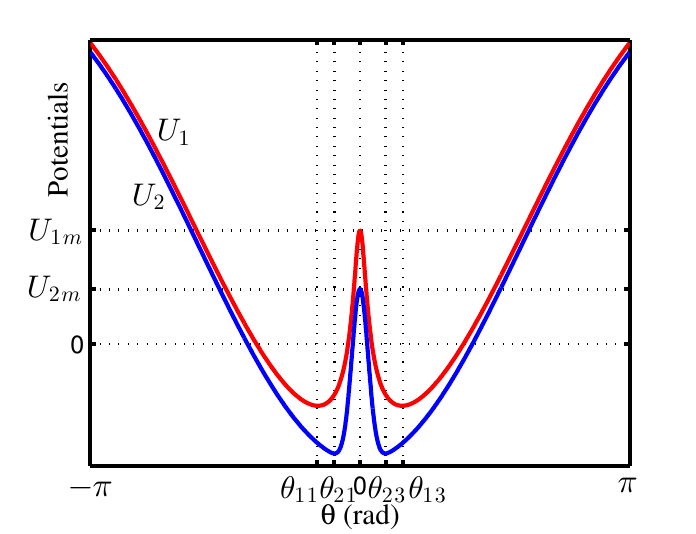}\\
		\hspace*{1.2cm} a) \hspace*{6.0cm} b) \hfill\\
		\caption{\label{FIGURE5}\footnotesize {Shape of the monostable (a) and the bistable (b) potential.}}
	\end{center}
\end{figure}

The expressions of the potentials derived from the first and second models are given, respectively, as
\begin{equation}\label{EQUAT15}
	U_1  = \frac{1}{2}C\theta ^2  - \gamma \cos \theta  + \frac{{2\sigma I}}{{\sqrt {1 + \varepsilon  - \cos \theta } }},
\end{equation}
\begin{equation}\label{EQUAT16}
	U_2  = \frac{1}{2}C\theta ^2  - \gamma \cos \theta  + aI\exp \left( { - \frac{{\theta ^2 }}{b}} \right) .
\end{equation}
The mathematical analysis of the two potentials shows that the models have a simple mono-stable potential for $I\leq 0$ and present a bistable potential when $I>0$. For illustration, the shapes of mono-stable potentials are shown in Figure \ref{FIGURE5}a while the shapes of bistable potentials are presented in Figure \ref{FIGURE5}b.		

The parameters $U_{1m}$ and $U_{2m}$ labeled on the graphs represent the potential barriers and are defined as
\begin{equation}\label{EQUAT17}
	U_{1m}  =  - \gamma  + \frac{{2\sigma I}}{{\sqrt \varepsilon  }}\;\;{\text{and}}\;\;U_{2m}  =  - \gamma  + aI.
\end{equation}

As shown, the potential barriers are linear functions of the current $I$. Because of the qualitative similarities between the results obtained from the two models, one can use the magnetic charges interaction formulation to analyze other dynamical structures with four control parameters which are the ratio $\sigma$, the current amplitude $I$, the current frequency, and the current duty cycle.

\section{Dynamical states when the coil is powered by a sinusoidal current}

$\quad$ In this subsection, a sinusoidal current source in the form $i = I_m \sin \left( {2\pi ft} \right)
$ is used to power the coil.

\subsection{Mathematical analysis}

\begin{figure}[h!]
	\begin{center}
		\includegraphics[width=5.2cm,height=4.2cm]{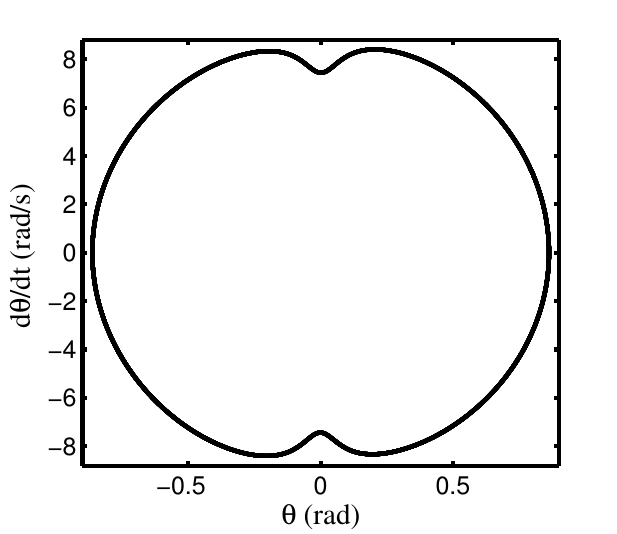}\hspace*{0.5cm}
		\includegraphics[width=6.8cm,height=4.2cm]{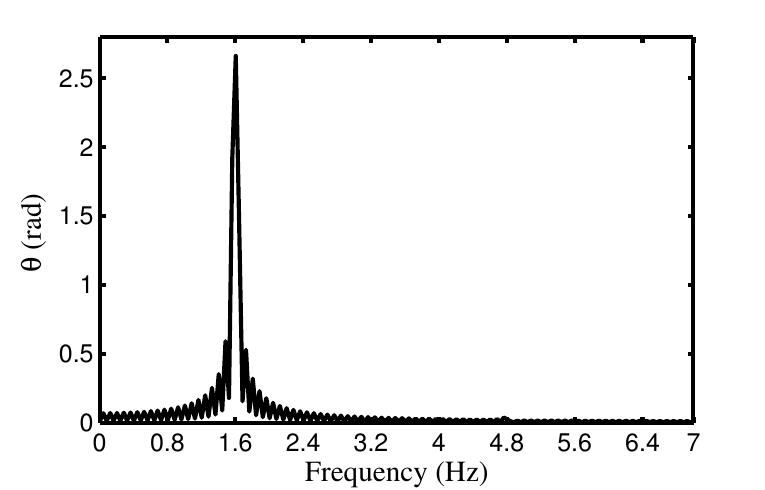}\\
		\hspace*{0.0cm} a) \hspace*{6.0cm} b) \hfill\\
		\caption{\label{FIGURE6A}\footnotesize {(a) Phase portrait of the system obtained for $f=3.2$ Hz and $I_m=200$ mA. (b) Corresponding frequency spectrum of angular displacement.}}
	\end{center}
\end{figure}

$\quad$ Numerical simulations of the nonlinear pendulum are carried out in order to evaluate the capability of the balance harmonic method to provide a correct approximated analytical solution of the differential equations given in the previous section. Figure \ref{FIGURE6A}a shows the phase portrait $\left( {\theta ,\;\dot \theta } \right)$ and Figure \ref{FIGURE6A}b shows the frequency spectrum of the angular displacement $\theta$ for $f=3.2$ Hz and $I_m=200$ mA.

\begin{figure}[h!]
	\begin{center}
		\includegraphics[width=5.2cm,height=4.2cm]{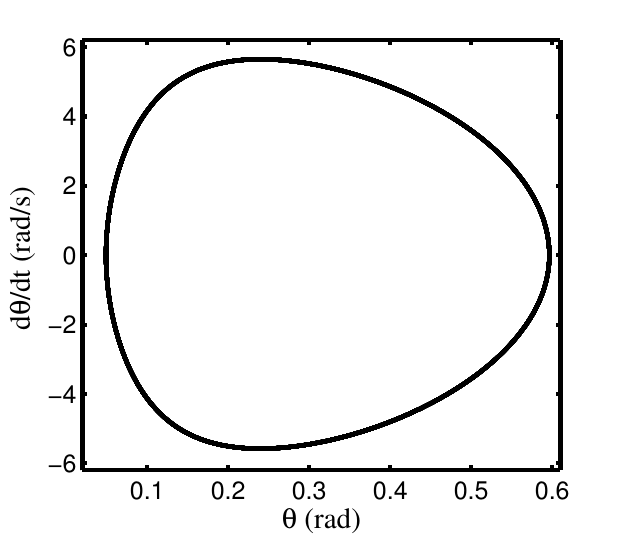}\hspace*{0.5cm}
		\includegraphics[width=6.8cm,height=4.2cm]{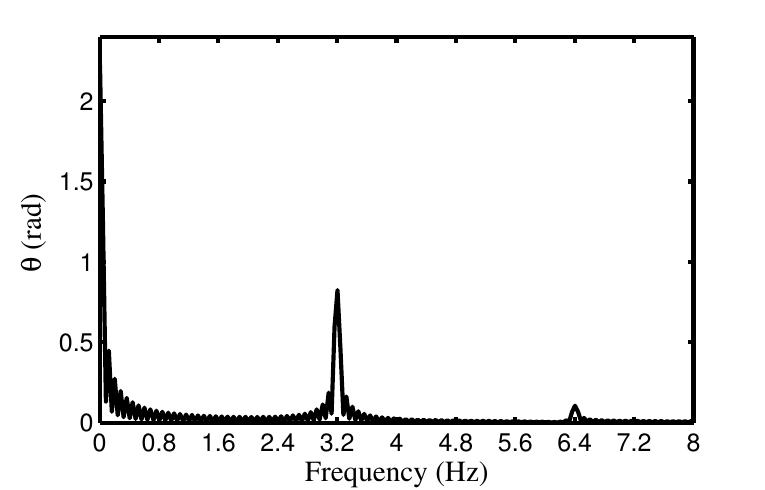}\\
		\hspace*{0.0cm} a) \hspace*{6.0cm} b) \hfill\\
		\caption{\label{FIGURE7A}\footnotesize {(a) Phase portrait of the system obtained for $f=3.2$ Hz and $I_m=500$ mA. (b) Corresponding frequency spectrum of angular displacement.}}
	\end{center}
\end{figure}

As shown in Figure \ref{FIGURE6A}, the pendulum oscillates with half of the excitation frequency  ($1.6$ Hz). A different situation is obtained when the amplitude of the excitation current is $I_m=500$ mA as shown in the graph of Figure \ref{FIGURE7A} also potted for $f=3.2$ Hz.

From the above dynamical behaviors of the angular movement of the pendulum, we can see a nonlinear interaction between the magnetic pendulum and the current-carrying coil through the external current source $i(t)$, evidenced by the peaks that occurred in the frequency spectra. Depending on the magnitude $I_m$ and frequency $f$ of the excitation current, the system can oscillate with frequencies $nf$, or with frequencies $ \left( {n - \frac{1}{2}} \right)f $ where $n$ is a positive integer. During the investigations, we have observed that the system provides one-side oscillations when it oscillates with frequencies integer multiple of $f$. We have also noticed that the system mostly oscillates with the frequencies half-integer multiple of $f$.

\begin{figure}[h!]
	\begin{center}
		\includegraphics[width=6.5cm,height=4.6cm]{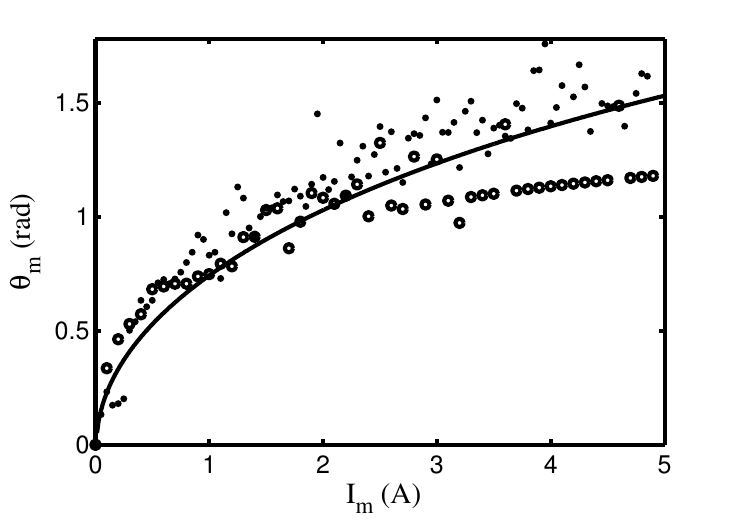}\hspace*{0.0cm}
		\includegraphics[width=6.5cm,height=4.6cm]{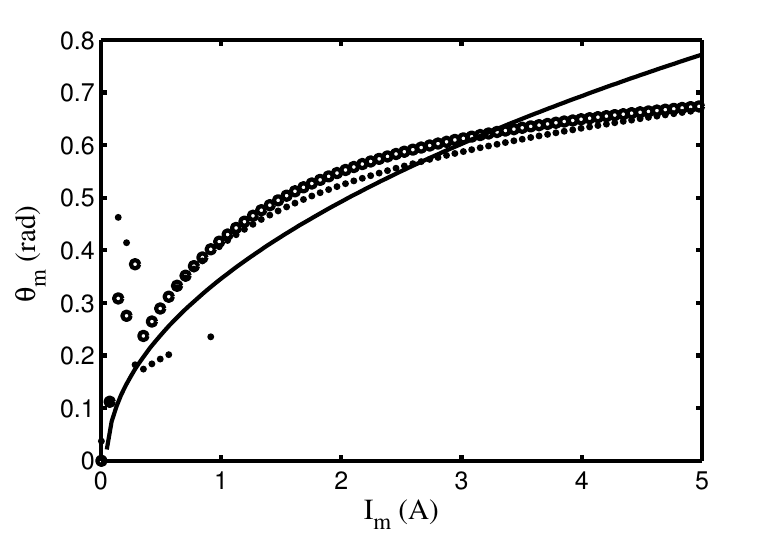}\\
		\hspace*{0.5cm} a) \hspace*{6.0cm} b) \hfill\\
		\caption{\label{FIGURE6}\footnotesize {(a) Amplitude of angle $\theta_m$ versus the magnitude $I_m$ of the current source for $f=1.1$ Hz. (b) Amplitude of angle $\theta_m$ versus the magnitude $I_m$ of the current source for $f=3.2$ Hz. The curves with the solid line are our analytical results, while the curves with points, and the curves with circles are obtained numerically from the first and second models, respectively.}}
	\end{center}
\end{figure}

The linearized equation of motion (\ref{EQUAT4}) is similar to the classical damped Mathieu equation. But to increase the precision of the analytical result, equation (\ref{EQUAT4}) is expanded up to the third order of the state variable $\theta$ as follows:
\begin{equation}\label{EQUAT18}
	\left( {J\ddot \theta  + \left( {\beta  + \chi M_s } \right)\dot \theta  + C\theta } \right)\left( {\varepsilon  + \frac{3}{4}\theta ^2 } \right) + \gamma \left( {\varepsilon \theta  + \frac{{9 - 2\varepsilon }}{{12}}\theta ^3 } \right) = \frac{{\sigma I_m }}{{\sqrt \varepsilon  }}\left( {\theta  - \frac{1}{6}\theta ^3 } \right)\sin \left( {\omega t} \right).
\end{equation}
Using an averaging method, the analytical solutions of equation (\ref{EQUAT18}) can be sought for time-periodic solutions in the form
\begin{equation}\label{EQUAT19}
	\theta  = A\cos \left( {\frac{{\omega t}}{2}} \right) + B\sin \left( {\frac{{\omega t}}{2}} \right),
\end{equation}
where $\theta _m  = \sqrt {A^2  + B^2 }$ is the amplitude of the analytical signal and represents our unknown variable here. This expression is known as the principal parametric resonance, and it corresponds to the subharmonic resonance since the forcing frequency is twice the frequency of the response signal. The substitution of the trigonometric equation (\ref{EQUAT19}) into the differential equation (\ref{EQUAT18}), and after some mathematical manipulations, we found that $\theta_m$ is the root of the following fourth-order polynomial equation
\begin{equation}\label{EQUAT20}
	\begin{gathered}
		A_2 \theta _m^4  - 2A_1 \theta _m^2  + A_0  = 0,\;\;{\text{where}} \hfill \\
		A_2  = 4\sigma ^2 I_m^2  - 81\varepsilon \left( {\beta  + \chi M_s } \right)^2 \omega ^2  - 9\varepsilon \left( {9J\omega ^2  + 2\varepsilon \gamma  - 9C - 9\gamma } \right)^2 , \hfill \\
		A_1  = 48\left( {\sigma ^2 I_m^2  + 9\varepsilon ^2 \left( {\beta  + \chi M_s } \right)^2 \omega ^2  + 3\varepsilon ^2 \left( {J\omega ^2  - C - \gamma } \right)\left( {9J\omega ^2  + 2\varepsilon \gamma  - 9C - 9\gamma } \right)} \right), \hfill \\  A_0  = 576\left( {\sigma ^2 I_m^2  - 4\varepsilon ^3 \left[ {\left( {\beta  + \chi M_s } \right)^2 \omega ^2  + \left( {J\omega ^2  - C - \gamma } \right)^2 } \right]} \right). \hfill \\
	\end{gathered}
\end{equation}
The possible positive roots of equation (\ref{EQUAT20}) are
\begin{equation}\label{EQUAT21}
	\theta _{1m}  = \sqrt {\frac{{A_1  - \sqrt {A_1^2  - A_0 A_2 } }}{{A_2 }}} \;{\text{ or }}\;\theta _{2m}  = \sqrt {\frac{{A_1  + \sqrt {A_1^2  - A_0 A_2 } }}{{A_2 }}}.
\end{equation}
To verify the result obtained analytically, the amplitude $\theta_m$ of the response is plotted as a function of the magnitude $I_m$ of the input current. Figure \ref{FIGURE6}a,b are plotted for the frequencies at $f=1.1$ Hz and $f=3.2$ Hz, respectively.

For both graphs, the analytical solutions are plotted with the full line, while the simulation results are plotted with points and with circles using the first and second models respectively. For the first value of the frequency $f=1.1$ Hz, the mechanical displacement of the pendulum has a chaotic behavior while it has approximatively a sinusoidal shape for the second value of the frequency $f=3.2$ Hz. One can notice a good agreement between the analytical and simulated results even in the case of chaotic dynamics as the simulated results fluctuate around the analytical ones. Also notice that, for the values of the parameters used in this work, only $\theta_{1m}$ given by equation (\ref{EQUAT21}) has positive values.

\begin{figure}[h!]
	\begin{center}
		\includegraphics[width=6.0cm,height=4.6cm]{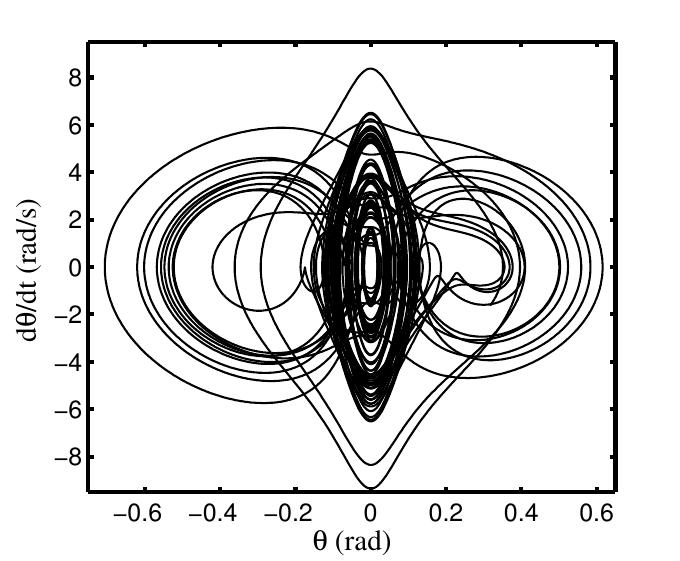}\hspace*{0.5cm}
		\includegraphics[width=6.0cm,height=4.6cm]{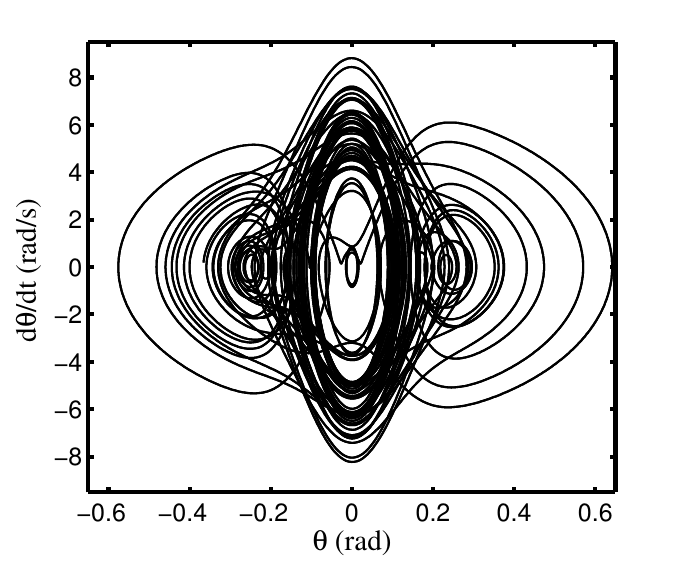}\\
		\hspace*{0.5cm} a) \hspace*{6.0cm} b) \hfill\\
		\caption{\label{FIGURE7}\footnotesize {Phase portraits of the pendulum computed for $f=1.1$ Hz and $I_m=550$ mA. Obtained using the first model (a) and the second model (b).}}
	\end{center}
\end{figure}

As one can see, the pendulum under the effect of the magnetic torque can behave both chaotically and periodically, depending on the parameters $f$ and $I_m$ of the external current source. The examples of chaotic motions are shown in Figure \ref{FIGURE7} for $I_m =550$ mA and $f = 1.1$ Hz.

\begin{figure}[h!]
	\begin{center}
		\includegraphics[width=6.0cm,height=4.6cm]{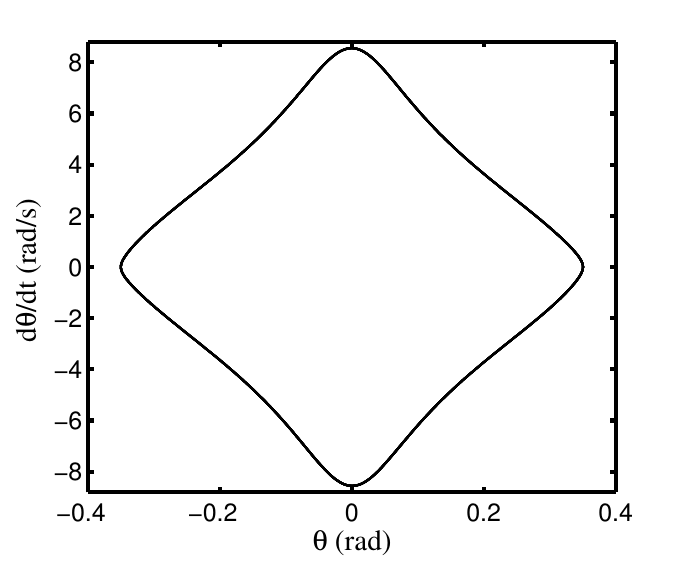}\hspace*{0.5cm}
		\includegraphics[width=6.0cm,height=4.6cm]{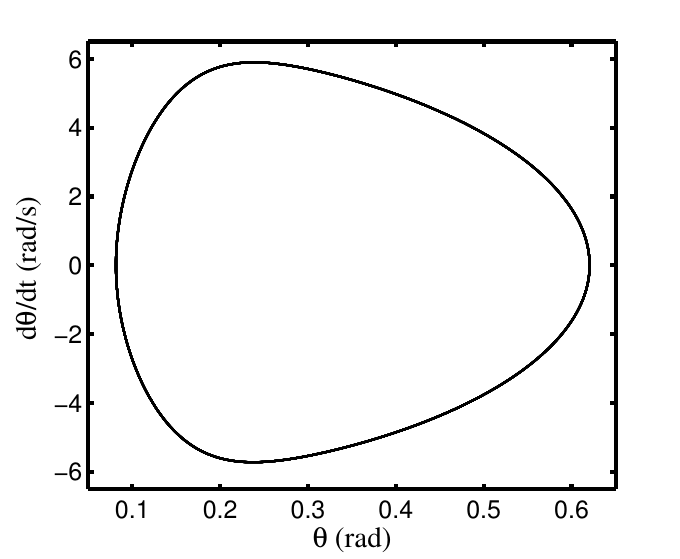}\\
		\hspace*{0.5cm} a) \hspace*{6.0cm} b) \hfill\\
		\caption{\label{FIGURE8}\footnotesize {Phase portraits of the pendulum computed for $f=3.2$ Hz and $I_m=550$ mA. Obtained using the first model (a) and the second model (b).}}
	\end{center}
\end{figure}

Figure \ref{FIGURE7}a is obtained with the first model and Figure \ref{FIGURE7}b is obtained using the second model. Even in this case, both models are showing similar irregular behaviors as indicated in Figure \ref{FIGURE6}. Apart from chaotic behavior, the pendulum can also exhibit periodic dynamics. \textcolor{black}{The regular rotations of the pendulum are presented in Figure \ref{FIGURE8} for $I_m = 550$ mA and $f = 3.2$ Hz}.

Figures \ref{FIGURE8}a,b are obtained using the first and second models, respectively. Both figures are showing periodic behavior and we have found that the system moves with half the frequency of the external excitation as predicted analytically.

\subsection{Bifurcation analysis}

\begin{figure}[h!]
	\begin{center}
		\includegraphics[width=6.5cm,height=5.0cm]{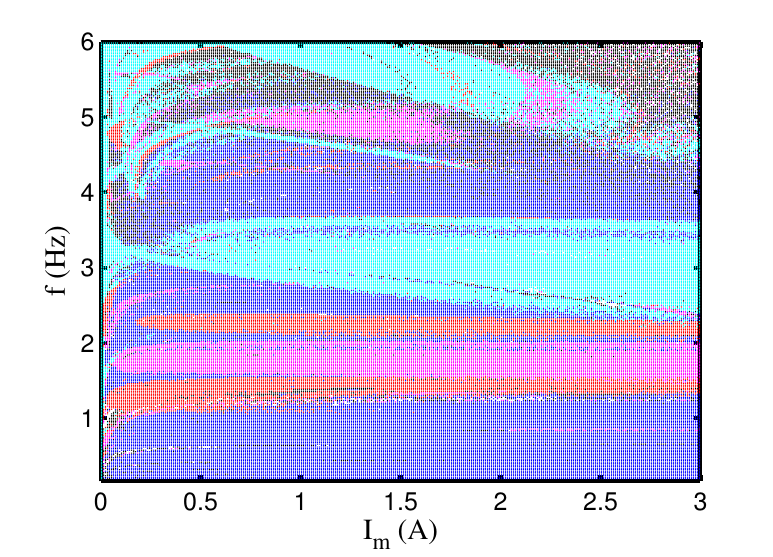}\hspace*{0.0cm}
		\includegraphics[width=6.5cm,height=5.0cm]{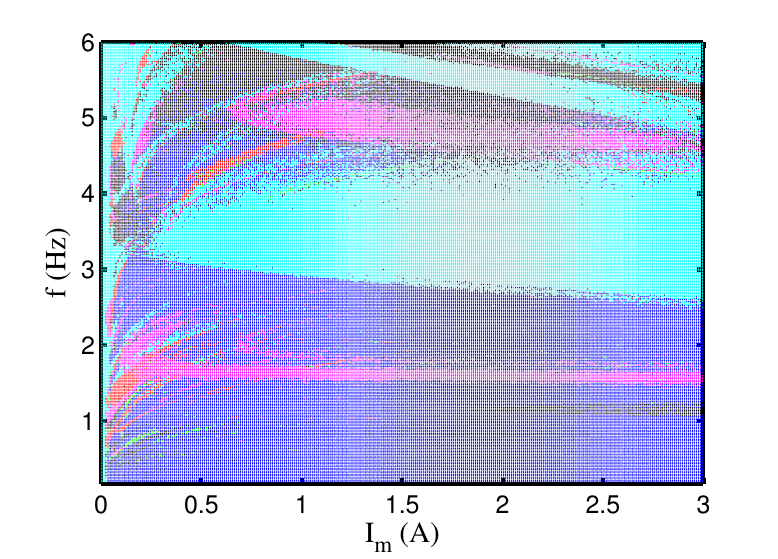}\\
		\hspace*{0.0cm} a) \hspace*{6.0cm} b) \hfill\\
		\caption{\label{FIGURE16}\footnotesize {Two-parameter bifurcation diagram showing the dynamical behavior of the system in the plane $\left( {f,\;I_m } \right)$ with a sinusoidal current source. Obtained using the first model (a) and the second model (b).}}
	\end{center}
\end{figure}

$\quad$ Since the systems described by the differential equations (\ref{EQUAT14}) and (\ref{EQUAT15}) have two control parameters $I_m$ and $f$ which are accessible experimentally, the different dynamical behaviors exhibited by the system can be best summarized by analyzing its bifurcation diagrams. Figure \ref{FIGURE16} shows the two-parameter bifurcation diagrams highlighting the dependence of the system on the two control parameters $f$ and $I_m$.

To identify the dynamical behavior of the system, the common way is to identify the number $n$ in the period$-n$ response when varying the control parameters. 
We have assigned a specific color to each period$-n$ response.
 Then the two-parameter color diagrams are created in Figures \ref{FIGURE16}a,b using the first and second models, respectively. In both figures, the cyan area, the red area, the magenta area, and the green areas denote the period-1, period-2, period-3, and period-4 behaviors of the system, respectively. Secondly, the black area indicates a combination of period-5, period-6, period-7, period-8, period-9, and period-10. Finally, all responses with $n$ greater than $10$ represent the chaotic dynamics and are indicated by the blue area.

One can notice a good similitude between the two bifurcation diagrams. Nevertheless, a significant region of period-2 is obtained with the first model while the chaotic behavior occupies more space with the second model.

\section{Dynamical states in presence a square periodic excitation}

$\quad$ In this section, the current-carrying coil is supplied with a square current source defined as $i\left( t \right) = \frac{{I_m }}{2}\left[ {1 + {\text{sign}}\left( {\sin \left( {2\pi ft} \right)} \right)} \right]$. For the harmonic balance method to work, the square current source $i(t)$ is to be represented by a corresponding Fourier series expansions $\tilde i\left( t \right)$, and we limit ourselves to the first harmonic expressed as
\begin{equation}\label{EQUAT122}
	\tilde i\left( t \right) = \frac{{2I_m }}{\pi }\left[ {\frac{\pi }{4} + \sin \left( {2\pi ft} \right)} \right].
\end{equation}

\subsection{Mathematical analysis}

$\quad$ With the Fourier series expansions of the input current given in equation (\ref{EQUAT122}), we will proceed as done in the previous section. The third order series expansion of equation (\ref{EQUAT14}) according to the variable $\theta$ has the form
\begin{equation}\label{EQUAT123}
	\left( {J\ddot \theta  + \left( {\beta  + \chi M_s } \right)\dot \theta  + C\theta } \right)\left( {\varepsilon  + \frac{3}{4}\theta ^2 } \right) + \gamma \left( {\varepsilon \theta  + \frac{{9 - 2\varepsilon }}{{12}}\theta ^3 } \right) = \eta \left( {\theta  - \frac{1}{6}\theta ^3 } \right)\left( {\phi  + \sin \left( {\omega t} \right)} \right),
\end{equation}
where the newly introduced parameters are defined as $\eta = \frac{{2\sigma I_m }}{{\pi \sqrt \varepsilon  }}\,{\text{ and }}\,\phi  = \frac{\pi }{4}$. Because of the additional constant term $\phi$ on the external force,  the analytical solution of equation (\ref{EQUAT123}) will have the following form
\begin{equation}\label{EQUAT124}
	\theta  =\theta_0 +A\cos \left( {\frac{\omega t}{2}} \right) + B\sin \left( {\frac{\omega t}{2}} \right).
\end{equation}
Terms $\theta_0$ and $\theta _m  = \sqrt {A^2  + B^2 }$ are the average and the amplitude of the analytical signal, respectively, and represent our unknown variables. The substitution of equation (\ref{EQUAT124}) into the differential equation (\ref{EQUAT123}), and after some mathematical manipulations, we found that the amplitude $\theta_m$ and the average $\theta_0$ are the solutions of the following fourth-order and third-order polynomial equations, respectively:
\begin{equation}\label{EQUAT125}
	A_2 \theta _m^4  - 2A_1 \theta _m^2  + A_0  = 0,
\end{equation}
\begin{equation}\label{EQUAT125A}
	\left( {B_2 \theta _m^2  + B_1 } \right)\theta _0^3  - B_0 \theta _{\text{0}}  = 0.
\end{equation}
The coefficients of equation (\ref{EQUAT125}) are defined as
\begin{equation}\label{EQUAT125B}
	\begin{gathered}
		A_2  = 4\eta \left( {1 - 9\phi ^2 } \right) + 36\eta \phi \left( {9\psi  + 2\varepsilon \gamma } \right) - 81\beta '^2 \omega ^2  - 9\left( {9\psi  + 2\varepsilon \gamma } \right)^2 , \hfill \\  A_1  = 48\left( {\eta ^2 \left( {1 - 6\phi ^2 } \right) + 3\eta \phi \left( {9\psi  + 2\varepsilon \gamma  - 2\varepsilon \psi } \right) + 9\varepsilon \beta '^2 \omega ^2  + 3\varepsilon \psi \left( {9\psi  + 2\varepsilon \gamma } \right)} \right), \hfill \\  A_0  = 576\left( {\eta ^2 \left( {1 - 4\phi ^2 } \right) - 8\eta \varepsilon \phi \psi  - 4\varepsilon ^2 \left[ {\beta '^2 \omega ^2  + \psi ^2 } \right]} \right), \hfill \\  B_2  = 18J\omega ^2  - 27\left( {C + \gamma } \right) + 6\varepsilon \gamma  - 6\eta \phi , \hfill \\  B_1  = -24\left( {\varepsilon C + \varepsilon \gamma  - \eta \phi } \right)\;\;{\text{and}}\;B_0  = 18\left( {C + \gamma } \right) - 4\varepsilon \gamma  + 4\eta \phi . \hfill \\\end{gathered}
\end{equation}
The constants $\beta'$ and $\psi$ are used to simplify the notations and are defined as $\beta ' = \beta  + \chi M_s $ and $\psi  = J\omega ^2  - C - \gamma$. The possible positive roots of equation (\ref{EQUAT125}) are expressed as
\begin{equation}\label{EQUAT126}
	\theta _{1m}  = \sqrt {\frac{{A_1  - \sqrt {A_1^2  - A_0 A_2 } }}{{A_2 }}} \;{\text{ or }}\;\theta _{2m}  = \sqrt {\frac{{A_1  + \sqrt {A_1^2  - A_0 A_2 } }}{{A_2 }}}.
\end{equation}
On the other hand, the average value $\theta_0$ is expressed as a function of the amplitude $\theta_m$ in equation (\ref{EQUAT127}).
\begin{equation}\label{EQUAT127}
	\theta _0  = 0\;\;{\text{or}}\;\;\theta _0  =  \pm \sqrt {\frac{{B_0 }}{{B_2 \theta _m^2  + B_1 }}}.
\end{equation}
If the expression under the square root of equation (\ref{EQUAT127}) is negative, the average term is zero. This corresponds to symmetric oscillations around $\theta=0$ rad for the case of a single stable equilibrium point. Otherwise, the expression with the square root has to be considered. The positive or negative signs are related to the initial conditions used during the numerical simulations of the differential equation.

\begin{figure}[h!]
	\begin{center}
		\includegraphics[width=6.5cm,height=4.5cm]{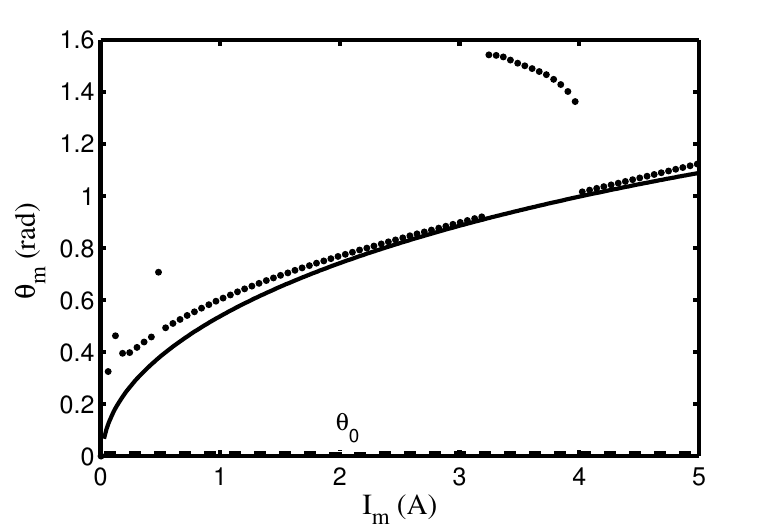}\hspace*{0.0cm}
		\includegraphics[width=6.5cm,height=4.5cm]{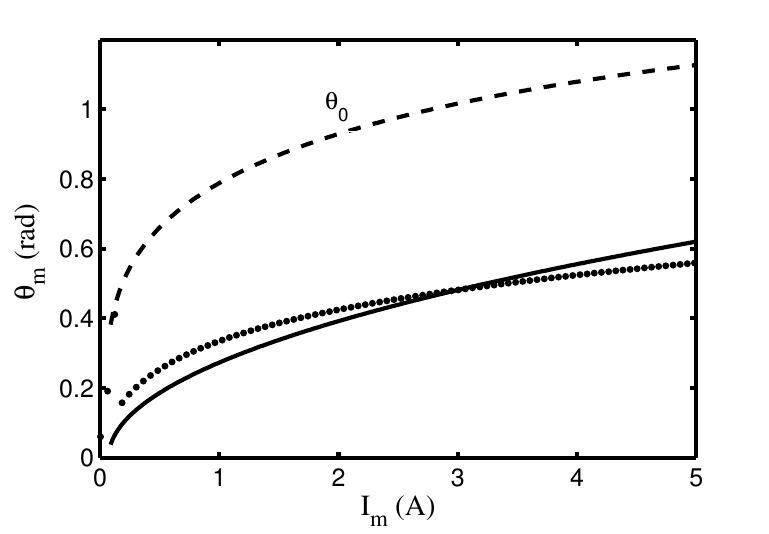}\\
		\hspace*{0.5cm} a) \hspace*{6.0cm} b) \hfill\\
		\caption{\label{FIGURE19}\footnotesize {(a) Amplitude of angle $\theta_m$ and average value $\theta_0$ versus the magnitude $I_m$ of the current source for $f=1.3$ Hz. (b) Amplitude of angle $\theta_m$ and average value $\theta_0$ versus the magnitude $I_m$ of the current source for $f=3.2$ Hz. The curves with the solid line and with dashed lines are our analytical results and represent the magnitude $\theta_m$ and average $\theta_0$ of the angular displacement, respectively. The curves with dots are obtained numerically from our first model and represent the magnitude $\theta_m$.}}
	\end{center}
\end{figure}

For illustration, we use the magnitude $I_m$ of the current source as the control parameter, and the system is analyzed for two different frequencies of the current source. Figures \ref{FIGURE19}a,b are plotted for $f=1.3$ Hz and $f=3.2$ Hz, respectively.

\begin{figure}[h!]
	\begin{center}
		\includegraphics[width=6.0cm,height=5.0cm]{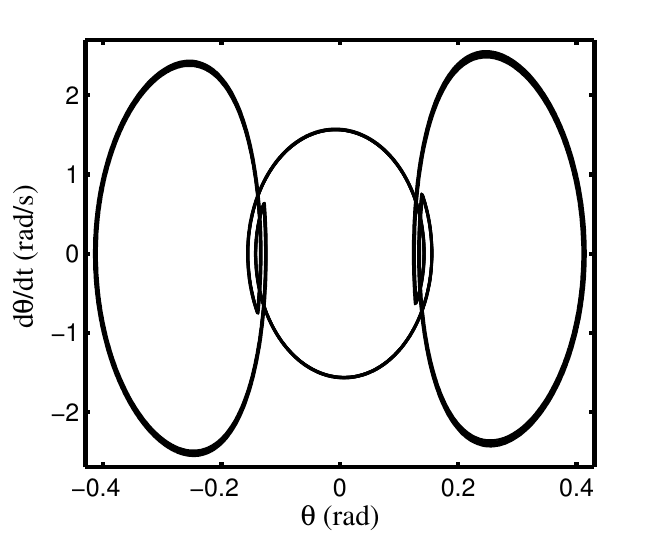}\hspace*{0.0cm}
		\includegraphics[width=6.6cm,height=5.0cm]{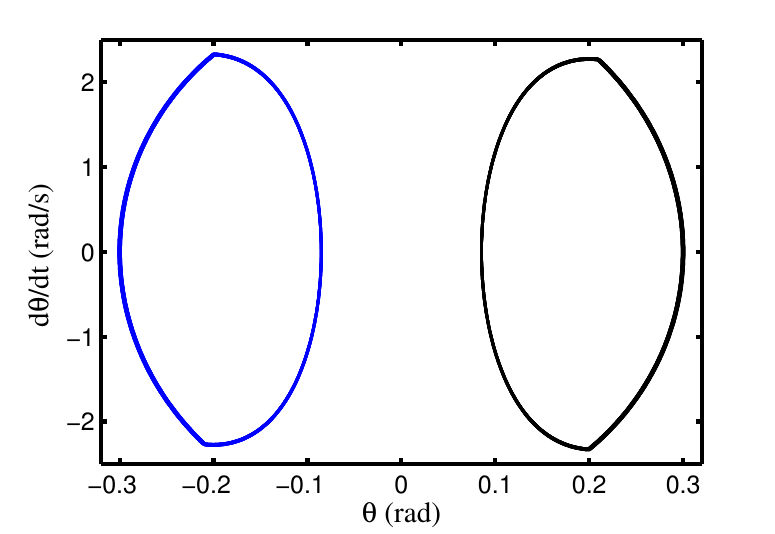}\\
		\hspace*{0.0cm} a) \hspace*{6.0cm} b) \hfill\\
		\caption{\label{FIGURE20}\footnotesize {(a) Phase portrait of the system obtained for $f=1.3$ Hz and $I_m=300$ mA. (b) Phase portrait of the system computed for $f=3.2$ Hz and $I_m=170$ mA and a different initial conditions. The black color curve corresponds to the initial condition ($\theta(0)=-0.3$ rad, $\dot{\theta}(0)=0.5$ rad/s), while the blue color curve corresponds to the initial condition ($\theta(0)=0.3$ rad, $\dot{\theta}(0)=-0.5$ rad/s).}}
	\end{center}
\end{figure}

The analytical solutions are plotted with the full line ($\theta_m$) and dashed lines ($\theta_0$), while the simulation results showing the magnitudes of the angular displacement are plotted with points. The simulation results of the average angular displacement (not represented here) are in very good agreement with the analytical results. As shown in the graphs, a good agreement is also found here between our analytical and simulation results. A slight disagreement observed between the analytical and numerical results can appear because just the first harmonic of the series Fourier expansion of the square current source has been considered in the analytical treatment. We have to notice that, in the case of equation (\ref{EQUAT125}), which has no positive real root, so one has to replace $\phi$ in the coefficients $A_n$ by zero. Also, higher harmonics have to be taken into consideration when looking at the solution of the differential equation. In these cases, the Fourier spectra will be formed by many harmonic components having basic, super-harmonic, sub-harmonic, and combination frequencies. Nevertheless, other characteristics of the system can be captured by the obtained analytical expressions. In fact, for $f=1.3$ Hz, the average angle given by equation (\ref{EQUAT127}) is $\theta_0=0$ rad. This is clearly visible in the phase portrait of Figure \ref{FIGURE20}a plotted for $f=1.3$ Hz and $I_m=300$ mA.
As predicted by the analytical result, the graph shows a symmetrical behavior around the origin point, indicating that the average rotational angle is zero ($\theta_0=0$ rad). The situation is quite different for $f=3.2$ Hz as highlighted on the graphs of Figure \ref{FIGURE20}b. The two curves are plotted for the same value of the current $I_m=170$ mA but different initial conditions. The curve in black is obtained for ($\theta(0)=-0.3$ rad, $\dot{\theta}(0)=0.5$ rad/s) while the curve in blue line is plotted for ($\theta(0)=0.3$ rad, $\dot{\theta}(0)=-0.5$ rad/s). As predicted in Figure \ref{FIGURE20}b and verified by the simulated results, the curves are centered at $\theta_0=\pm 0.235$ rad. This is an important result since equation (\ref{EQUAT127}) can be used to determine the condition to be satisfied by the parameters to obtain side oscillations.

\subsection{Bifurcation analysis}

\begin{figure}[h!]
	\begin{center}
		\includegraphics[width=6.5cm,height=5.0cm]{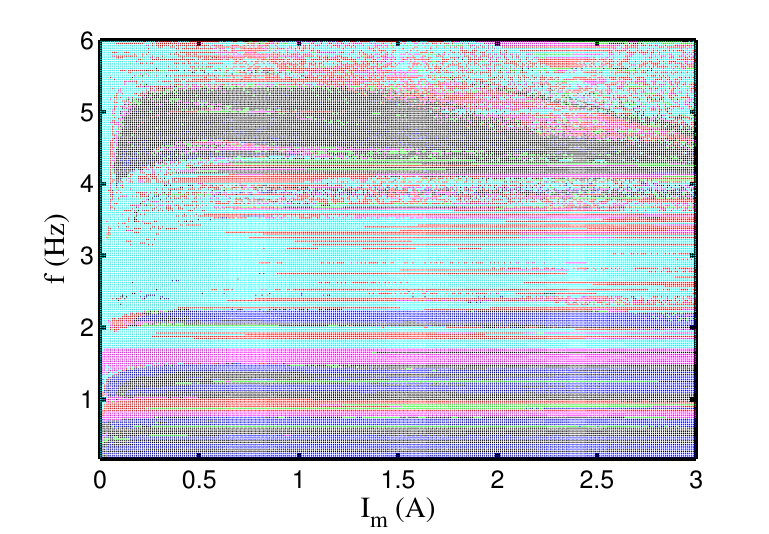}\hspace*{0.0cm}
		\includegraphics[width=6.5cm,height=5.0cm]{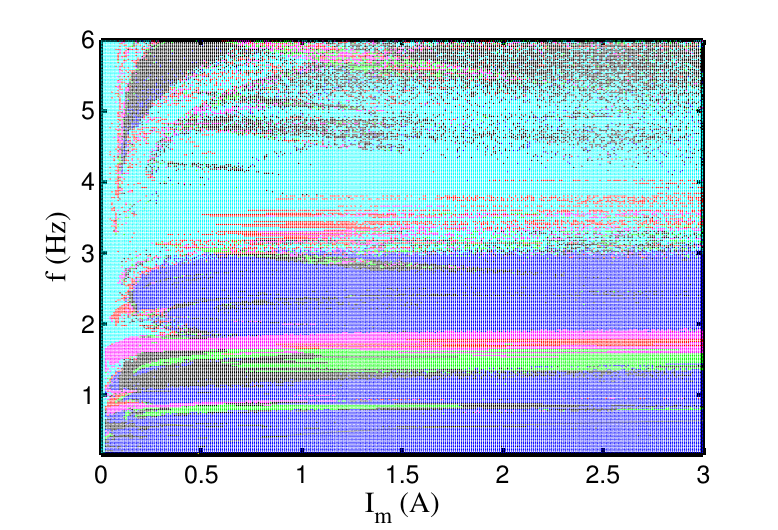}\\
		\hspace*{0.0cm} a) \hspace*{6.0cm} b) \hfill\\
		\caption{\label{FIGURE17}\footnotesize {Two-parameter bifurcation diagram showing the dynamical behavior of the system in the plane $\left( {f,\;I_m } \right)$ with a square current source. Obtained using the first model (a) and the second model (b).}}
	\end{center}
\end{figure}

$\quad$ Proceeding as above, Figure \ref{FIGURE17} shows the two-parameter bifurcation diagrams of the system using the two control parameters $f$ and $I_m$.

Figure \ref{FIGURE17}a is obtained using the first model, while Figure \ref{FIGURE17}b is plotted using the second model. The two bifurcation diagrams show many similar qualitative results. The color configuration is maintained as in the previous section: the cyan area, the red area, the magenta area, and the green areas denote the period-1, period-2, period-3, and period-4 behaviors of the system, respectively. The black area indicates a combination of  period-5, period-6, period-7, period-8, period-9, and period-10 while the chaotic dynamics are indicated by the blue area. It is evident that the character of the system's motion strongly depends both on the frequency and amplitude of the excitation.

\section{Experiment}
In this section, the experimental rig is described and comparisons between experimental and simulation results are presented.
\subsection{Experimental setup}

\begin{figure}[h!]
	\begin{center}
		\includegraphics[width=11.0cm,height=5.0cm]{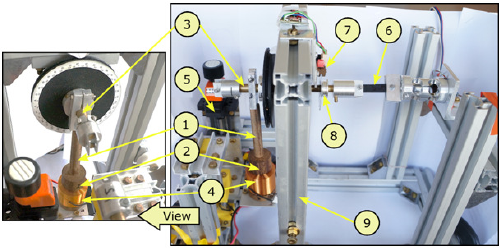}\hspace*{0.0cm}
		\caption{\label{FIGURE21A}\footnotesize {Experimental setup: (1)- pendulum, (2)- neodymium magnet, (3)- brass axis, (4)- electric coil, (5)- linear lift, (6)- an elastic element, (7)- optical sensor, (8)- code wheel, (9)- aluminium frame.}}
	\end{center}
\end{figure}

${\quad}$ Figure \ref{FIGURE21A} shows the experimental setup that was used for the experiment.
The setup consists of a magnetic pendulum (1) equipped with a neodymium magnet (2), mounted on a~brass axis (3). The neodymium magnet measures $22$ mm in diameter and $10$ mm in height. Pair of rolling bearings support the brass axis. Under the pendulum is a fixed platform containing an electric coil (4). Through a linear lift (5), the platform's vertical position can be controlled, adjusting the clearance between the coil and the pendulum at rest. A~clearance of $1.6$ mm was used in the presented studies. The coil properties are as follows: inductance $22$ mH, resistance $10.6\,\Omega$, copper wire diameter $0.5$ mm, external diameter $40$ mm, bore diameter 17 mm, height 10 mm. The fixed base is connected to the opposite end of the brass axis through a flexible rubber element (6). The purpose of this rubber element is to enhance the restoring torque within the analyzed system. The rubber element possesses a rectangular cross-section measuring $8.5 \times 6.5 $mm, with a length of $40.3$ mm.

\begin{figure}[h!]
	\begin{center}
		\includegraphics[width=7.0cm,height=4.0cm]{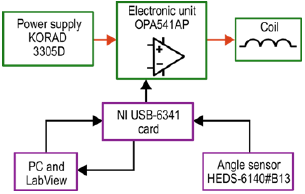}\hspace*{0.0cm}
		\caption{\label{FIGURE21B}\footnotesize {Signal-flow diagram, red arrows indicate current signals whereas black arrows indicate voltage signals.}}
	\end{center}
\end{figure}

Figure \ref{FIGURE21B} illustrates the signal flow interaction among the components within the studied system.

The laboratory power supply (KORAD KA3305D) energizes the electrical coil through a dedicated electronic module that utilizes an operational amplifier (OPA541AP). Operating in current source mode, the power supply ensures a consistent current output, unaffected by alterations in the coil wire resistance due to thermal effects. The coil current signal waveform tracks the voltage signal provided to the electronic unit via an NI USB-6341 card. LabView software controls the voltage signal originating from the NI USB-6341 card.

The electric coil current signal stimulates the creation of a magnetic field enveloping the coil, which interacts with the magnet. Throughout the experiments, the presumption was made that a positive current signal induces magnetic repulsion from the coil, while a negative current brings about attraction toward the coil. Due to the thermal limitations of the coil and the amplifier, it was not possible to perform experiments in which the coil current would be higher than $2$ A, moreover, for the same reason, it was not possible to perform long-term experiments (e.g. bifurcation analysis) for currents greater than $1.5$ A.

The optical incremental sensor (HEDS-6140-B13), denoted as (7) and working with code wheel (8), captures the pendulum's angular position. The angular position sensor offers a resolution of $0.36$ degrees. Materials for constructing the setup, including frame (9), were chosen from non-magnetic alternatives, such as aluminum alloys, composite materials, or polymers.

\subsection{Comparison between the experimental and numerical results}

\begin{figure}[h!]
	\begin{center}
		\includegraphics[width=4.2cm,height=3.8cm]{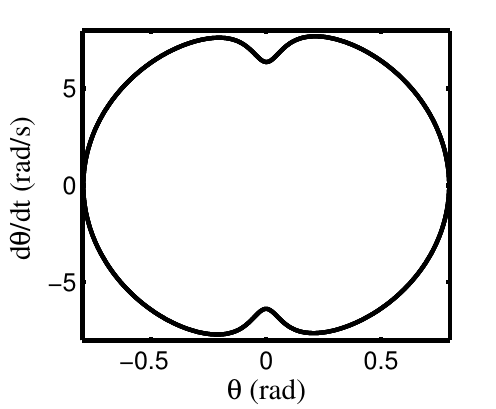}\hspace*{0.0cm}
		\includegraphics[width=4.2cm,height=3.8cm]{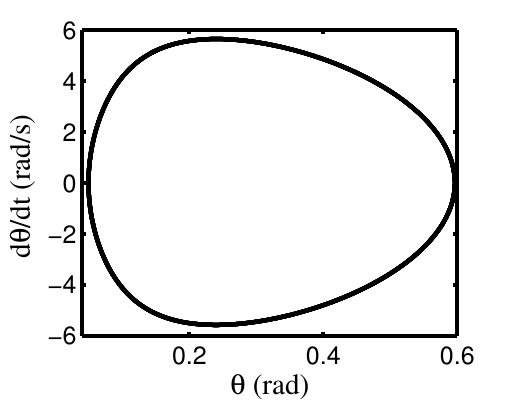}\hspace*{0.0cm}
		\includegraphics[width=4.2cm,height=3.8cm]{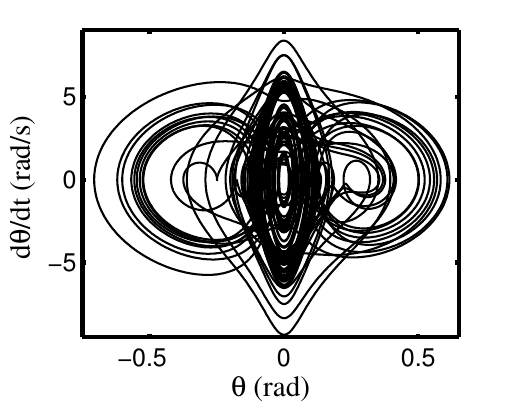}\\
		\hspace*{0.2cm} a) \hspace*{3.8cm} b) \hspace*{3.7cm} c)\hfill\\
		\includegraphics[width=4.2cm,height=3.8cm]{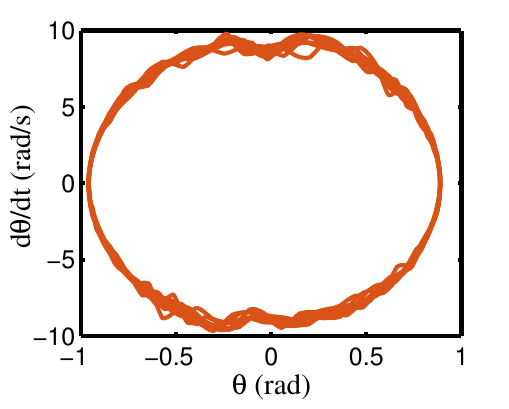}\hspace*{0.0cm}
		\includegraphics[width=4.2cm,height=3.8cm]{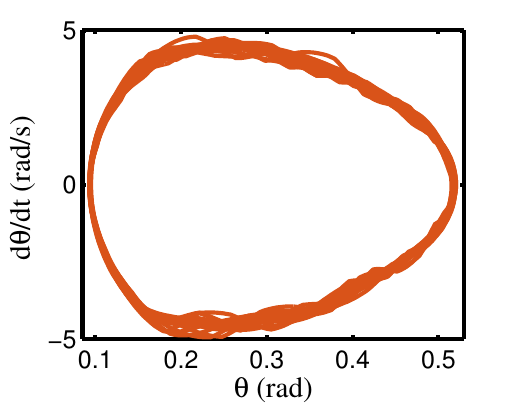}\hspace*{0.0cm}
		\includegraphics[width=4.2cm,height=3.8cm]{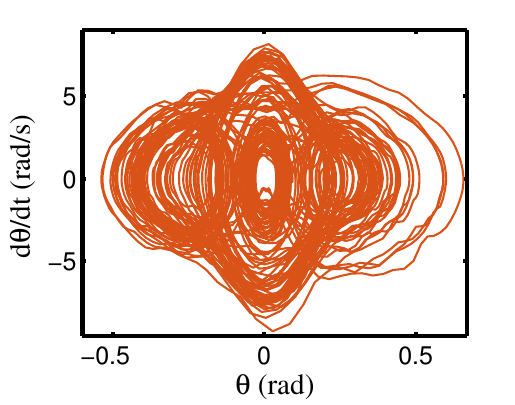}\\
		\hspace*{0.2cm} d) \hspace*{3.8cm} e) \hspace*{3.7cm} f)\hfill\\
		\includegraphics[width=4.2cm,height=3.8cm]{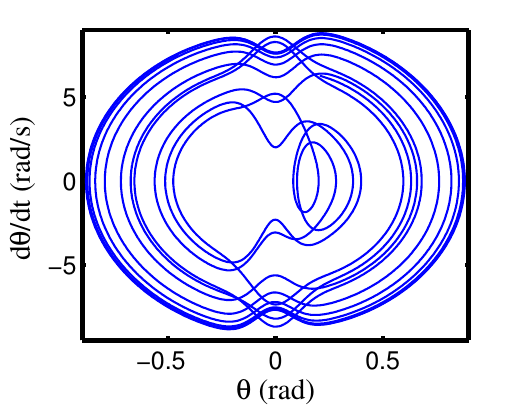}\hspace*{0.0cm}
		\includegraphics[width=4.2cm,height=3.8cm]{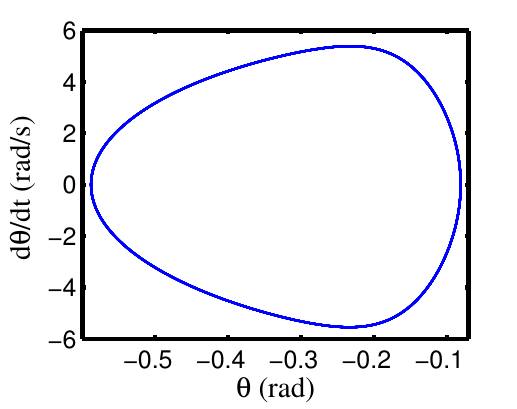}\hspace*{0.0cm}
		\includegraphics[width=4.2cm,height=3.8cm]{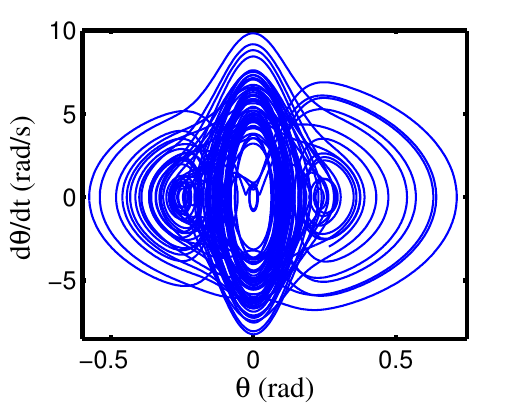}\\
		\hspace*{0.2cm} \textcolor{black}{g)} \hspace*{3.8cm} \textcolor{black}{h)} \hspace*{3.7cm} \textcolor{black}{i)}\hfill\\
		\caption{\label{FIGURE22} Different phase portraits of the system. The black and blue curves are obtained numerically using the first and second models, respectively. The brown curves are obtained experimentally. System has the following parameters:
		($I_m=230$ mA, $f=3.2$ Hz) for (a), (d), (g);  ($I_m=500$ mA, $f=3.2$ Hz) for (b), (c), (h); ($I_m=550$ mA, $f=1.1$ Hz) for (c), (f), (i).}
	\end{center}
\end{figure}

${\quad}$ In this section, the experimental results are compared to the simulation ones. We first consider the case where the coil is supplied by a sinusoidal current source. For different values of the amplitudes and frequencies of the current, some phase portraits were obtained numerically and experimentally and are presented in Figure \ref{FIGURE22}.

Figures \ref{FIGURE22}a-c are calculated from direct numerical simulations of the first model (\ref{EQUAT4}) for set of the following parameters: ($I_m=230$ mA, $f=3.2$ Hz), ($I_m=500$ mA, $f=3.2$ Hz) and ($I_m=550$ mA, $f=1.1$ Hz), respectively. To verify these results, the corresponding experimental results are plotted in Figures \ref{FIGURE22}d-f, respectively. Similarly, Figures \ref{FIGURE22}g-i are obtained from direct numerical simulations of the second model (\ref{EQUAT5}). The experimental investigations of the systems showed that it also exhibits regular and irregular behaviors. While there is good agreement in the overall trends in the phase portraits, there are small differences between the experimental and the simulation results.

\begin{figure}[h!]
	\begin{center}
		\includegraphics[width=6.5cm,height=4.5cm]{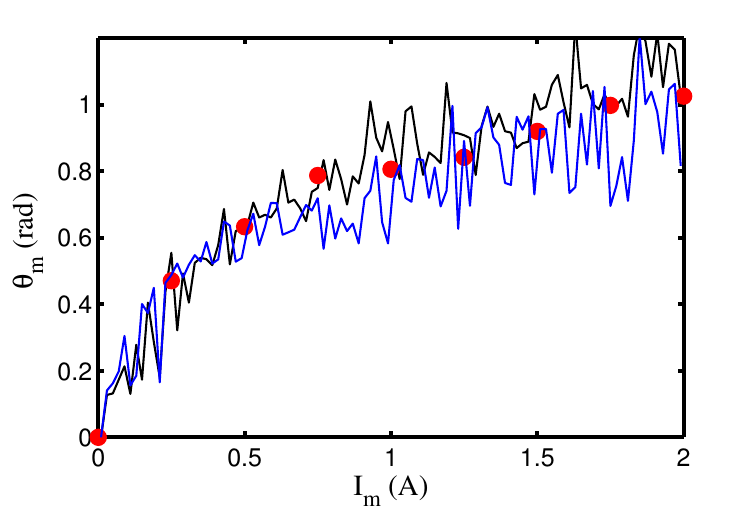}\hspace*{0.0cm}
		\includegraphics[width=6.5cm,height=4.5cm]{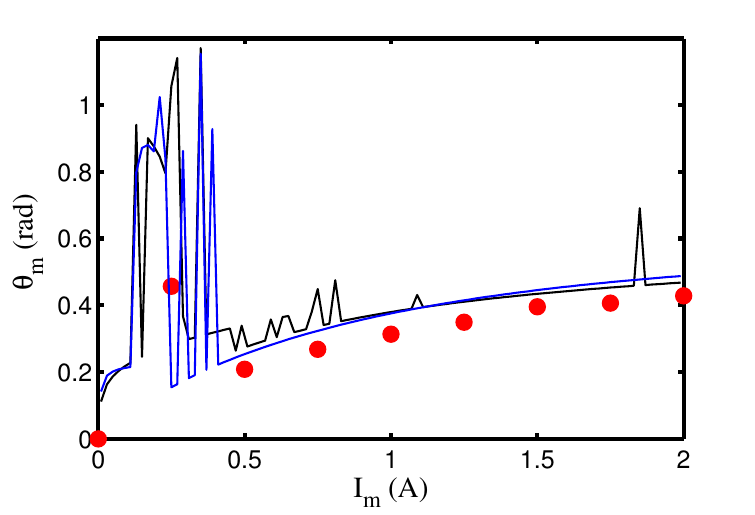}\\
		\hspace*{0.0cm} a) \hspace*{6.0cm} b) \hfill\\
		\caption{\label{FIGURE23}\footnotesize {Amplitude of angle $\theta_m$ versus the magnitude $I_m$ of the current source computed for current frequencies: (a) $f=1.1$ Hz and (b) $f=3.2$ Hz. The black and blue curves are our simulated results obtained from the first and second models, respectively, while the  points are obtained experimentally.}}
	\end{center}
\end{figure}

To further test, the models described in this work, the experimental amplitude responses of the magnetic pendulum are generated and compared to the simulations ones. The amplitude responses for the physical pendulum were generated in the experiment by increasing the magnitude of the current source from $0.0$ A to $2.0$ A with a step size of $0.5$ A. Simulated amplitude response diagrams were generated in these ranges, but with a finer step size of $0.02$ A. Both experimental and simulated results can be seen in Figure \ref{FIGURE23}.

The curves in Figure \ref{FIGURE23}a,b are plotted for $f=1.1$ Hz and $f=3.2$ Hz, respectively. As shown in the graphs, the magnetic pendulum is in the chaotic state for $f=1.1$ Hz and the simulation results from both models fluctuate around the experimental results. For the frequency equals $3.2$ Hz, the amplitudes of the system recorded experimentally are slightly less than those provided by the simulations. Nevertheless, excellent qualitative agreement is found between the simulated results and the experimental ones.

\begin{figure}[h!]
	\begin{center}
		\includegraphics[width=5.5cm,height=4.2cm]{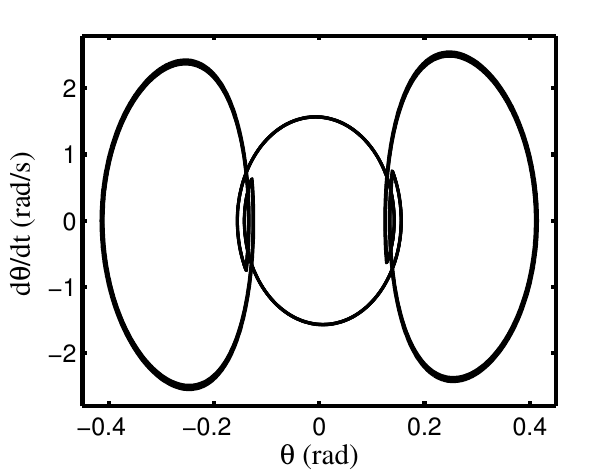}\hspace*{0.0cm}
		\includegraphics[width=6.5cm,height=4.2cm]{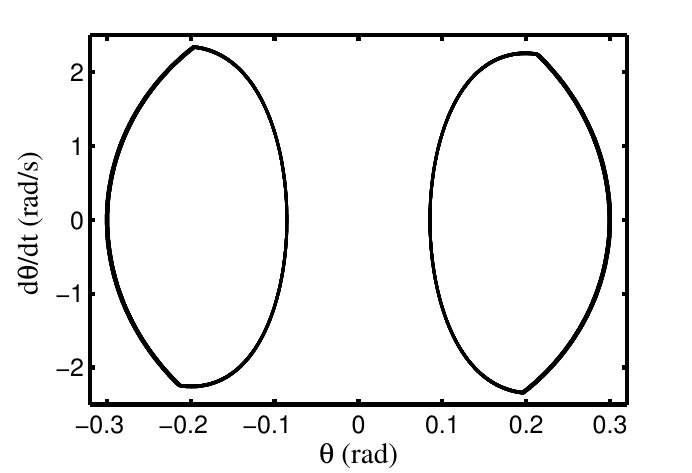}\\
		\hspace*{0.0cm} a) \hspace*{5.4cm} b) \hfill\\
		\includegraphics[width=5.5cm,height=4.2cm]{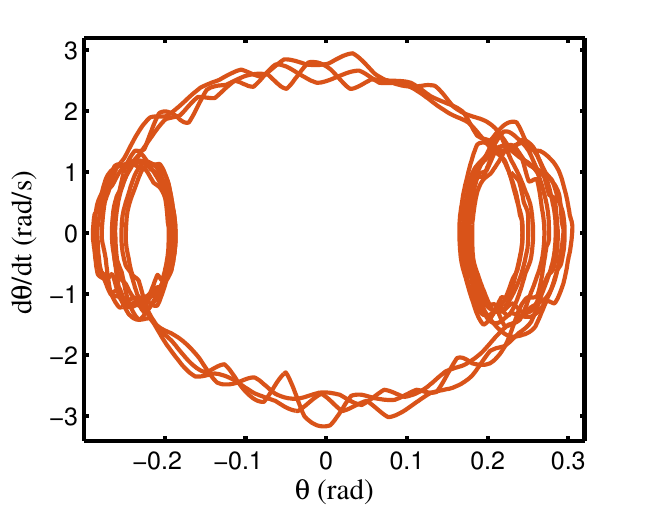}\hspace*{0.0cm}
		\includegraphics[width=6.5cm,height=4.2cm]{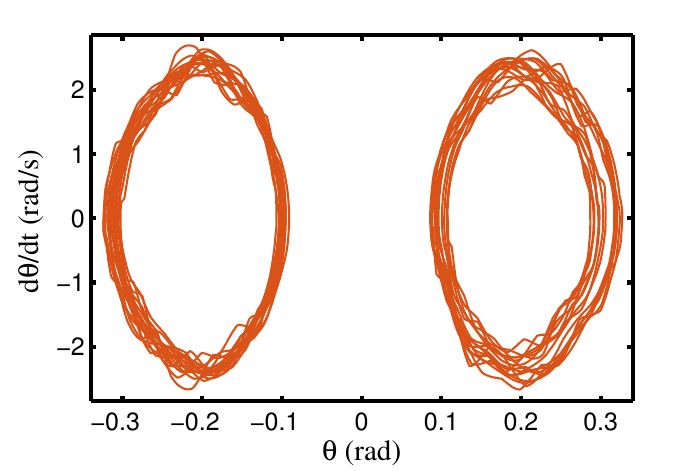}\\
		\hspace*{0.0cm} c) \hspace*{5.4cm} d) \hfill\\
		\includegraphics[width=5.5cm,height=4.2cm]{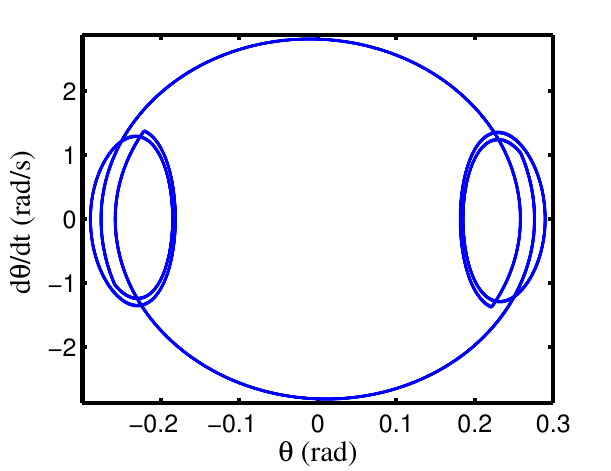}\hspace*{0.0cm}
		\includegraphics[width=6.5cm,height=4.2cm]{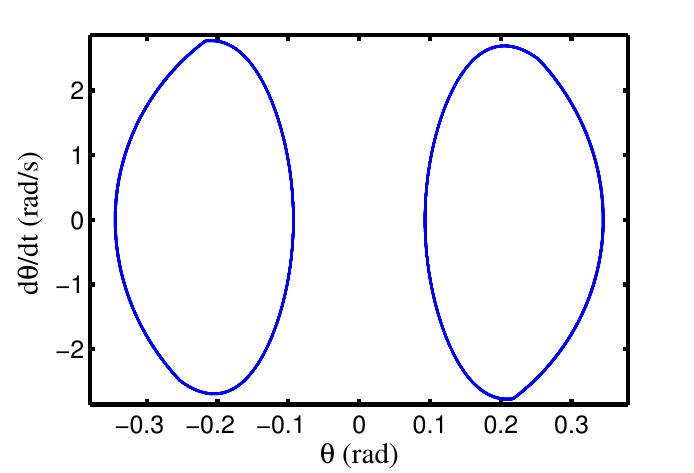}\\
		\hspace*{0.0cm} e) \hspace*{5.4cm} f) \hfill\\
		\caption{\label{FIGURE24} Different phase portraits of the system. The black and blue curves are obtained numerically using the first and second models, respectively. The brown curves are obtained experimentally. System has the following parameters: ($I_m=300$ mA, $f=1.3$ Hz) for (a), (c), (e); ($I_m=170$ mA, $f=3.2$ Hz) for (b), (d), (f) .}
	\end{center}
\end{figure}

Finally, the case where the coil magnet is supplied by a square current source is investigated. Proceeding as above, some phase portraits obtained numerically and experimentally are presented in Figure \ref{FIGURE24} for different values of the amplitudes and frequencies of the source.

Figures \ref{FIGURE24}a,b are obtained numerically for ($I_m=300$ mA, $f=1.3$ Hz) and ($I_m=170$ mA, $f=3.2$ Hz)  using the first and second models, respectively. The corresponding experimental results are plotted in Figures \ref{FIGURE24}c,d, respectively. Figures \ref{FIGURE24}e,f are also obtained numerically using the second model described above. The multistability in the system is well predicted by both models and confirmed by the experimental investigations as illustrated in Figures \ref{FIGURE24}b,d,f. The curves located in the negative $\theta-$axis are obtained by using the initial conditions $\theta(0)=0.3$ rad and $\dot \theta \left( 0 \right)=-0.5$ rad/s whereas the curves located in the positive $\theta-$axis are obtained by using the initial conditions $\theta(0)=-0.3$ rad, $\dot \theta \left( 0 \right)=0.5$ rad/s. Even here, both mathematical models of the system showed good agreement with the experimental results.

\begin{figure}[h!]
	\begin{center}
		\includegraphics[width=6.5cm,height=4.5cm]{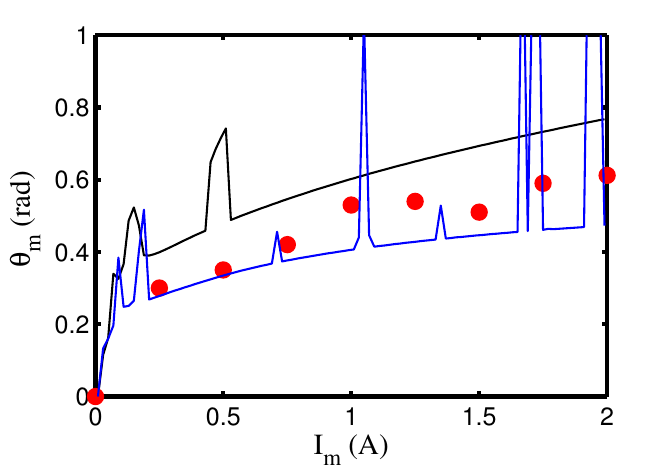}\hspace*{0.0cm}
		\includegraphics[width=6.5cm,height=4.5cm]{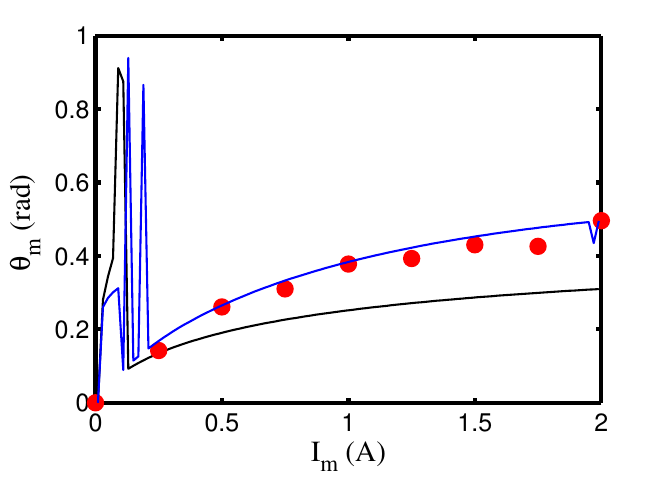}\\
		\hspace*{0.0cm} a) \hspace*{6.0cm} b) \hfill\\
		\caption{\label{FIGURE25}\footnotesize {			
				Amplitude of angle $\theta_m$ versus the magnitude $I_m$ of the current source computed for current frequencies: (a) $f=1.3$ Hz and (b) $f=3.2$ Hz. The black and blue curves are our simulated results obtained from the first and second models, respectively, while points are obtained experimentally.}}
	\end{center}
\end{figure}

Finally, we keep the frequency fixed and vary the magnitude of the current source to explore the dynamical behavior of the system forced by a square current source. Both the simulation and experimental results are presented in Figure \ref{FIGURE25}a,b plotted for $f=1.3$ Hz and $f=3.2$ Hz, respectively.

In both figures, the black and blue curves are our simulated results obtained from the first and second models, respectively, while the  points are obtained experimentally. As one can notice, the experimental results lie between the simulated ones but in this case, are more closed to the results obtained with the experimentally fitted interaction model.

\section{Conclusion}

${\quad}$ In this paper, the theoretical and experimental studies of a magnetic pendulum are presented. First, the corresponding mathematical models using two different approaches have been derived. The coil is powered by a sinusoidal current source and a square current source. Based on the harmonic balance method, the approximate analytical solutions of the amplitude of the mechanical displacement of the pendulum have been found analytically and verified through numerical simulations. Very rich dynamical states are obtained when varying the control parameters. To more visualize the system behavior, two-parameter bifurcation diagrams have been plotted considering the current amplitude and frequency as the control parameters. Good agreements are found between our theoretical and experimental results. It has been found that the two approaches of approximating the magnetic interaction give results very close.

In this study, to simplify our analysis, the neodymium magnet and the coil were considered to be point magnets. It would be interesting to consider the physical parameters of the magnets. This will allow to analyze the effects of the number of turns and the dimensions of the coil on the system.

\section*{Acknowledgments}
${\quad}$ Paul Woafo acknowledges the support from the Ulam Program of the Polish National Agency for Academic Exchange (grant n. BPBN/ULM/2022/1/00056/Dec/1). The authors thank the support of the Polish National Science Centre, Poland under the Grant OPUS 18 No. 2019/35/B/ ST8/00980. Moreover, the work of Krystian Polczyński (experimental set-up modifications, experimental measurements, computational data processing) was supported by the National Science Center, Poland under the grant PRELUDIUM 20 No. 2021/41/N/ST8/01019. \textcolor{red}{For the purpose of Open Access, the authors has applied a CC-BY public copyright licence to any Author Accepted Manuscript (AAM) version arising from this submission. This is an AAM of an article published by Elsevier in Mechanical Systems and Signal Processing on 15 January 2024, available at: https://doi.org/10.1016/j.ymssp.2024.111114}

\bibliographystyle{elsarticle-num}
\bibliography{mybibfile}
\end{document}